\documentclass[9pt,conference]{IEEEtran}

\usepackage{graphicx}
\usepackage{amssymb}
\usepackage{graphicx}
\usepackage{amsmath}
\usepackage{color}
\usepackage{dsfont}
\usepackage{amsthm}

\def\Bin{{\sf Bin}}
\def\expect{{\mathbb E}}
\def\E{{\mathbb E}}

\def\cF{{\cal F}}

\def\cS{{\cal S}}

\def\E{\mathbb{E}}

\def\de{{\rm d}}
\def\ind{{\mathbb I}}

\def\sQ{{\Box}}
\def\sC{{\pm}}
\def\sT{{+}}
\def\sN{{\sf N}}

\def\de{{\rm d}}

\def\reals{{\mathbb R}}
\def\E{{\mathbb E}}
\def\<{\langle}
\def\>{\rangle}

\def\prob{{\mathbb P}}

\def\sign{{\rm sign}}

\def\CG{\mbox{\tiny\rm CG}}
\def\SE{\mbox{\tiny\rm SE}}
\def\LS{\mbox{\tiny\rm LS}}
\def\pb{{\chi}}
\def\MSE{{\sf MSE}}
\def\MM{\mbox{\tiny\rm MM}}
\def\ind{{\mathbb I}}

\newcommand{\gly}{{\chi}}

\newcommand{\bR}{{\bf R}}
\newcommand{\calF}{{\cal F}}
\newcommand{\goto}{\rightarrow}
\newcommand{\eps}{\epsilon}
\newcommand{\bitem}{\begin{itemize}}
\newcommand{\eitem}{\end{itemize}}
\newcommand{\beq}{\begin{eqnarray}}
\newcommand{\eeq}{\end{eqnarray}}
\newtheorem{finding}{Finding}
\newtheorem{theorem}{Theorem}[section]

\newtheorem{definition}[theorem]{Definition}
\newtheorem{proposition}[theorem]{Proposition}

\title{Message Passing Algorithms\\
for Compressed Sensing}

\author{\authorblockN{David L. Donoho }
\authorblockA{Department of Statististics\\
Stanford University\\
 donoho@stanford.edu
  } \and
  \authorblockN{ Arian Maleki}
\authorblockA{Department of Electrical Engineering\\
Stanford University\\
 arianm@stanford.edu
  } \and
  \authorblockN{Andrea Montanari }
\authorblockA{Department Electrical Engineering\\
and Department of Statistics\\
Stanford University\\
 montanar@stanford.edu
  }

  }

\begin{document}




\maketitle


\begin{abstract}
Compressed sensing aims to undersample certain
high-dimensional signals, yet accurately reconstruct them by exploiting signal
characteristics.  Accurate reconstruction
is possible when the object to be recovered is
sufficiently sparse in a known basis.  Currently,
the best known sparsity-undersampling tradeoff
is achieved when reconstructing by convex optimization --
which is expensive in important large-scale applications.

Fast iterative thresholding algorithms have
been intensively studied as alternatives to
convex optimization for large-scale problems.
Unfortunately known fast algorithms
offer substantially worse sparsity-undersampling tradeoffs than
convex optimization.

We introduce a simple costless modification
to iterative thresholding
making the sparsity-undersampling tradeoff
of the new algorithms equivalent to that of the
corresponding convex optimization procedures.
The new iterative-thresholding algorithms are inspired
by belief propagation in graphical models.

Our empirical measurements of the sparsity-undersampling
tradeoff for the new algorithms agree with theoretical calculations.
We show that a state evolution formalism
correctly derives the true sparsity-undersampling tradeoff.
There is a surprising agreement between
earlier calculations based on random convex
polytopes and this new, apparently
very different theoretical formalism.
\end{abstract}

%
%
\section{Introduction and overview}

Compressed sensing
refers to a growing body of techniques that `undersample' high-dimensional
signals and yet recover them accurately  \cite{Donoho1,CaRoTa06}.
Such techniques make fewer measurements than traditional sampling theory demands:
rather than sampling proportional to frequency bandwidth, they make
only as many measurements as the underlying `information content'
of those signals. However, as compared with traditional sampling theory,
which can recover signals by applying simple linear reconstruction
formulas, the task of signal recovery from reduced measurements
requires nonlinear, and so far, relatively expensive reconstruction schemes.
One popular class of reconstruction schemes uses linear programming (LP) methods;
there is an elegant theory for such schemes  promising large
improvements over ordinary sampling rules in recovering sparse signals.
However, solving the
required LPs is substantially more expensive in applications than
the linear reconstruction schemes that are now standard.  In certain
imaging problems, the signal to be acquired may be an image with $10^6$
pixels and the required LP would involve tens of thousands of constraints
and millions of variables. Despite advances in the speed of LP,
such problems are still dramatically more expensive to solve
than we would like.

This paper develops an iterative algorithm
achieving reconstruction performance
 in one important sense {\it identical to} LP-based reconstruction
while  running dramatically faster.
We assume that a vector $y$
of $n$ measurements is obtained from
an unknown $N$-vector $x_0$ according to $y = Ax_0$,
where $A$ is the $n \times N$ measurement matrix $n < N$. Starting from
an initial guess $x^0 = 0$, the
\emph{first order approximate message passing}
(AMP) algorithm proceeds iteratively according to:
\begin{eqnarray}
x^{t+1} & = & \eta_t(A^*z^t+x^t)\, ,
\label{eq:FOAMP1}\\
z^t & = & y -Ax^t + \frac{1}{\delta}z^{t-1}\<\eta'_t(A^*z^{t-1}+x^{t-1})\>\, .
\label{eq:FOAMP2}
\end{eqnarray}
Here $\eta_t(\,\cdot\,)$ are scalar {\it threshold} functions (applied
componentwise), $x^t\in\reals^N$ is the current estimate of $x_0$,
and $z^{t}\in\reals^n$ is the current residual. $A^*$ denotes transpose of $A$.
For a vector $u=(u(1),\dots,u(N))$,
$\<u\> \equiv \sum_{i=1}^N u(i)/N$. Finally $\eta'_t(\,s\,) = \frac{\partial}{\partial s}\eta_t(\ s\,)$.

Iterative thresholding algorithms of other types have been popular
among researchers for some years,  the focus being on
schemes of the form
\begin{eqnarray}
x^{t+1} & = & \eta_t(A^*z^t+x^t)\, ,
\label{eq:IT1}\\
z^t & = & y -Ax^t .
\label{eq:IT2}
\end{eqnarray}
Such schemes can have very low per-iteration cost and low storage requirements;
they can attack very large scale applications, \--  much larger than
standard LP solvers can attack.  However, [\ref{eq:IT1}]-[\ref{eq:IT2}]
fall short of the sparsity-undersampling
tradeoff offered by LP reconstruction
 \cite{MaDo09sp}.

Iterative thresholding schemes based on [\ref{eq:IT1}], [\ref{eq:IT2}]
lack the crucial term in [\ref{eq:FOAMP2}] -- namely,
 $ \frac{1}{\delta}z^{t-1}\<\eta'_t(A^*z^{t-1}+x^{t-1})\>$
is not included.   We derive this term from
the theory of belief propagation in graphical models, and show that it
substantially improves the sparsity-undersampling tradeoff.

\begin{figure}[t]
\centerline{\includegraphics[width=1.\linewidth]{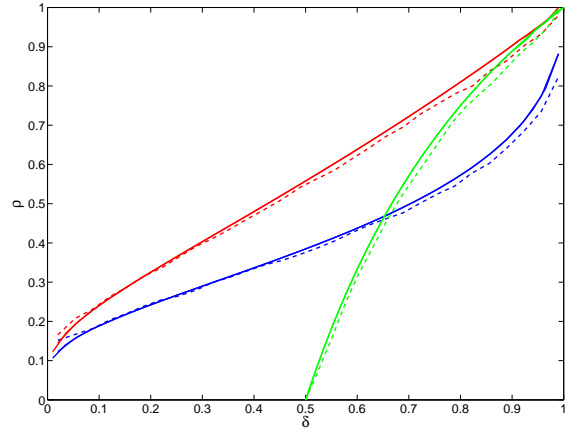}}
\caption{The phase transition lines for reconstructing
sparse non-negative vectors (problem $\sT$, red), sparse signed vectors
(problem $\sC$, blue) and vectors with entries in $[-1,1]$
(problem $\sQ$, green). Continuous lines refer to analytical predictions from
combinatorial geometry or the state
evolution formalisms. Dashed lines present data from experiments
with the AMP algorithm, with signal length $N=1000$ and $T=1000$ iterations.
For each value of $\delta$, we considered a grid of $\rho$ values,
at each value, generating $50$ random problems. The dashed line
presents the estimated 50th percentile of the response curve. At that percentile,
the root mean square error after $T$ iterations obeys
$\sigma_T\le 10^{-3}$ in half of the simulated reconstructions.}\label{fig:PhaseTrans}
\end{figure}

Extensive numerical
and Monte Carlo work reported here shows that
AMP, defined by eqns
[\ref{eq:FOAMP1}], [\ref{eq:FOAMP2}]  achieves a
sparsity-undersampling tradeoff matching the
theoretical tradeoff which has been proved for
LP-based reconstruction.  We consider a parameter space with axes quantifying
sparsity and undersampling.
In the limit of large dimensions $N,n$, the parameter space splits in two
{\it phases}:
one where the MP approach is  successful  in accurately reconstructing $x_0$
and one where it is unsuccessful. References \cite{Do05,DoTa05,DoTa08ArXiv} derived
regions of success and failure for LP-based recovery.
We find these two ostensibly different partitions of the
sparsity-undersampling parameter space to be {\it identical}.
Both reconstruction approaches succeed or fail over the same regions,
see Figure \ref{fig:PhaseTrans}.

Our finding has extensive empirical evidence
and strong theoretical support.
We introduce a {\it state evolution} formalism and find that it
accurately predicts the dynamical behavior of
numerous observables
of the AMP algorithm. In this formalism, the mean squared error of reconstruction
is a state variable; its change from iteration to iteration is modeled
by a simple scalar function, the  {\it MSE map}. When this map
has nonzero fixed points, the formalism predicts that
AMP will not successfully recover the desired solution.
The MSE map
depends on the underlying sparsity and undersampling ratios,
and can develop nonzero fixed points over a region
of sparsity/undersampling space.
The region is evaluated analytically
and found to coincide very precisely (ie. within numerical precision)
with the region over which LP-based methods are proved to fail.
Extensive Monte Carlo testing of AMP reconstruction
finds the region where AMP fails is, to within statistical
precision, the same region.

In short we introduce a fast iterative algorithm which is found
to perform as well as corresponding linear programming based methods
on random problems. Our findings are supported from simulations and from a
theoretical formalism.

Remarkably, the success/failure phases of LP
reconstruction were previously found by methods in combinatorial geometry;
we give here what amounts to a very simple formula for the phase
boundary, derived using a very different
and seemingly elegant theoretical principle.

%
%
\subsection{Underdetermined Linear Systems}

Let $x_0\in \reals^N$ be the signal of interest. We are
interested in reconstructing it from the vector of measurements
$y = Ax_0$, with $y\in\reals^n$, for $n<N$. For the moment, we assume
the entries $A_{ij}$ of the measurement matrix are independent and
identically distributed  normal $\sN(0,1/n)$.

We consider three canonical models for the signal $x_0$
and three nonlinear reconstruction procedures based
on linear programming.

\vspace{0.1cm}

\noindent $\sT$: $x_0$ is nonnegative, with at most
$k$ entries different from $0$.
Reconstruct by solving the LP: minimize $\sum_{i=1}^Nx_i$
subject to $x\ge 0$, and $Ax=y$.

\vspace{0.1cm}

\noindent $\sC$: $x_0$ has as many as $k$ nonzero entries.
Reconstruct by solving the  minimum $\ell_1$ norm problem: minimize $||x||_1$,
subject to $Ax=y$. This can be cast as an LP.

\vspace{0.1cm}

\noindent $\sQ$: $x_0 \in [-1,1]^N$, with at most $k$ entries in the interior
$(-1,1)$. Reconstruction by solving the LP feasibility problem: find any vector $x\in [-1,+1]^N$
with $Ax=y$.

\vspace{0.1cm}

\noindent  Despite the fact that the systems
are underdetermined, under certain conditions on $k,n,N$
these  procedures perfectly recover $x_0$.
This takes place subject to a {\it sparsity-undersampling tradeoff}
namely an upper bound on the signal complexity $k$
relative to $n$ and $N$.

%
%
\subsection{Phase Transitions}

The sparsity-undersampling tradeoff can most
easily be described by taking a large-system limit.
In that limit, we fix parameters $(\delta, \rho)$
in $(0,1)^2$ and let $k,n,N\to\infty$
with $ k/n \goto \rho $ and $n/N \goto \delta$.
%
The sparsity-undersampling behavior
we study is controlled by $(\delta, \rho)$,
with $\delta$ the undersampling fraction and
$\rho$ a measure of sparsity (with larger $\rho$ corresponding to more
complex signals).

The domain $(\delta,\rho) \in (0,1)^2$
has two phases, a `success' phase, where
exact reconstruction typically occurs,
and a `failure' phase were exact reconstruction
typically fails.  More formally, for each choice of $\pb \in \{\sT,\sC,\sQ\}$
there is a function $\rho_{\CG}(\cdot;\pb)$ whose graph partitions
the domain into two regions. In the `upper' region, where
$\rho > \rho_{\CG}(\delta; \pb)$, the corresponding LP reconstruction $x_1(\pb)$ fails
to recover $x_0$, in the following sense:
as $k,n,N \goto \infty$ in the large system limit with
$ k/n \goto \rho $ and $n/N \goto \delta$, the probability
of exact reconstruction $\{ x_1(\pb) = x_0\}$ tends to zero exponentially fast.
In the `lower' region, where
$\rho < \rho_{\CG}(\delta; \pb)$, LP reconstruction succeeds
to recover $x_0$, in the following sense:
as $k,n,N \goto \infty$ in the large system limit with
$ k/n \goto \rho $ and $n/N \goto \delta$, the probability
of exact reconstruction  $\{ x_1(\pb) = x_0\}$  tends to one exponentially fast.
We refer to \cite{Do05,DoTa05,DoTa08,DoTa08ArXiv}
for proofs and precise definitions of the curves
$\rho_{\CG}(\cdot;\pb)$.

The three functions
$\rho_{\CG}(\, \cdot\, ;\sT)$, $ \rho_{\CG}(\, \cdot\, ;\sC)$,
$\rho_{\CG}(\,\cdot\,;\sQ)$
are shown in Figure 1; they are the red, blue, and green curves, respectively.
The ordering $\rho_{\CG}(\delta;\sT)> \rho_{\CG}(\delta;\sC)$
(red $ >$ blue)
says that knowing that a signal is sparse and positive is more
valuable than only knowing it is sparse.
Both the red and blue curves
behave as $\rho_{\CG}(\delta;\sT,\sC)
\sim (2 \log(1/\delta))^{-1}$ as $\delta \to 0$;
surprisingly large amounts of undersampling are possible,
if sufficient sparsity is present.
 In contrast,
$\rho_{\CG}(\delta;\sQ)=0$
(green curve) for $\delta < 1/2$ so
the bounds $[-1,1]$ are really of no help unless we use
a limited amount of undersampling,
i.e. by less than a factor of two.

Explicit expressions for $\rho_{\CG}(\delta;\sT,\sC)$
are given in \cite{Do05,DoTa05}; they are quite involved
and use methods from combinatorial geometry.
By Finding 1 below, they agree to within numerical
precision to the following formula:
\beq \label{eq:rhose}
\rho_{\SE}(\delta;\pb)  =  \max_{z\ge 0}\left\{
\frac{1-(\kappa_{\pb}/\delta)\big[(1+z^2)\Phi(-z)-z\phi(z)\big]}{
1+z^2-\kappa_{\pb}\big[(1+z^2)\Phi(-z)-z\phi(z)\big]}\right\}\, ,
\eeq
where $\kappa_{\pb}=1$, $2$ respectively for $\pb=\sT$, $\sC$.
This formula, a principal result of this
paper, uses methods unrelated to combinatorial geometry.

%
\subsection{Iterative Approaches}

\, Mathematical results for the large-system limit
correspond well to application needs. Realistic
modern problems in spectroscopy and medical
imaging demand reconstructions of objects
with tens of thousands or even millions
of unknowns.  Extensive testing of practical
convex optimizers in these problems \cite{DoTa08_universality}
has shown that the large system asymptotic accurately describes
the observed behavior of computed solutions to the
above LPs.  But the same testing shows that
existing convex optimization algorithms run slowly
on these large problems, taking minutes or even hours
on the largest problems of interest.

Many researchers have abandoned
formal convex optimization, turning to fast iterative methods instead
\cite{HeGiTr06,Gilbert_OMP,IndykRuzic}.

The iteration [\ref{eq:FOAMP1}]-[\ref{eq:FOAMP2}]
is very attractive because it does not require the
solution of a system of linear equations, and because it does
not require explicit operations on the matrix $A$; it only requires
that one apply the operators $A$ and $A^*$ to any given vector.
In a number of applications \-- for example Magnetic Resonance Imaging \--
the operators $A$ which make practical sense are not really Gaussian random
matrices, but rather random sections of the Fourier transform and other
physically-inspired transforms \cite{CaRoTa06,LuDoSaPa08}.  Such operators can be applied
very rapidly using FFTs, rendering the above iteration extremely fast.
Provided the process stops after a limited
number of iterations,  the computations are very practical.

The thresholding functions $\{\eta_t(\,\cdot\,)\}_{t\ge 0}$ in these schemes
depend on both iteration and problem setting. In this paper
we consider $\eta_t(\,\cdot\, ) =
\eta( \cdot ; \lambda \sigma_t, \pb)$, where $\lambda$ is a threshold
control parameter,
$\pb \in \{ \sT, \sC, \sQ\}$ denotes the setting,
and $\sigma_t^2 ={\rm Ave}_j \E\{(x^t(j)-x_0(j))^2\}$ is the mean
square error of the current current estimate $x^t$ (in practice an
empirical estimate of this quantity is used).

For instance, in the case of sparse signed vectors (i.e. problem setting $\sC$),
we apply soft thresholding
$\eta_t(u) = \eta(u;\lambda \sigma,\sC)$, where
\begin{eqnarray}
\eta(u;\lambda\sigma,\sC) = \left\{
\begin{array}{ll}
(u-\lambda\sigma) & \mbox{ if $u\ge\lambda\sigma$,}\\
(u+\lambda\sigma) & \mbox{ if $u\le-\lambda\sigma$,}\\
0 & \mbox{ otherwise,}
\end{array}\right.\label{eq:SoftThresh}
\end{eqnarray}
where we dropped the argument $\sC$  to lighten notation.
Notice that $\eta_t$ depends
on the iteration number $t$ only through the mean square error (MSE)
$\sigma_t^2$.
%
%
\subsection{Heuristics for Iterative Approaches}

Why should the iterative approach work, i.e.
why should it converge to the correct answer $x_0$?
The case $\sC$
has been most discussed and we focus
on that case for this section.  Imagine first of all that $A$
is an orthogonal matrix, in particular
$A^* = A^{-1}$. Then the iteration [\ref{eq:FOAMP1}]-[\ref{eq:FOAMP2}] stops in 1 step,
correctly finding $x_0$. Next,
imagine that $A$ is an invertible
matrix;  \cite{DaDeDe04},
has shown that a related thresholding algorithm
with clever scaling of $A^*$
and clever choice of threshold, will correctly find $x_0$.
Of course both of these motivational observations assume $n=N$, so we are
not really {\it under}sampling.

We sketch a motivational argument for thresholding
in the truly undersampled case $n < N$ which is statistical,
which has been popular with engineers \cite{LuDoSaPa08}
and which leads to a proper `psychology' for understanding our results.
Consider the operator $H = A^*A - I$,
and note that  $A^* y =  x_0 + H x_0$.
If $A$ were orthogonal, we would of course have $H=0$,
and the iteration would, as we have seen immediately
succeed in one step.
If $A$ is a Gaussian random matrix and $n < N$, then
of course $A$ is not invertible and $A^*$ is not $A^{-1}$.
Instead of $Hx_0 = 0$, in the undersampled
case $H x_0$ behaves as a kind
of noisy random vector, i.e. $A^* y =  x_0 +{\sf noise}$.
Now $x_0$ is supposed to be a sparse vector,
and, one can see, the {\sf noise} term is
accurately modeled as a vector with i.i.d. Gaussian entries with
variance $n^{-1}\|x_0\|_2^2$.

In short, the first iteration gives us
a `noisy' version of the sparse vector we are seeking to recover.
The problem of recovering a sparse vector from
noisy measurements has been heavily discussed \cite{DJ94a}
and it is well understood that soft thresholding
can produce a reduction in mean-squared error
when sufficient sparsity is present and the
threshold is chosen appropriately.
Consequently, one anticipates that $x^1$ will
be closer to $x_0$ than $A^*y$.

At the second iteration, one has
$ A^*(y - Ax^1) = x_0 + H(x_0 - x^1)$.
Naively, the matrix $H$ does
not correlate with $x_0$ or $x^1$,
and so we might pretend
that $H(x_0-x^1)$ is again a Gaussian vector whose entries
have variance $n^{-1}||x_0-x^1||_2^2$.
This `noise level' is smaller than at iteration zero,
and so thresholding of this noise
can be anticipated to produce an even more accurate result
at iteration two; and so on.
%
%

There is a valuable digital communications
interpretation of this process. The vector $w = H x_0$
is the cross-channel interference or {\it mutual access
interference} (MAI), i.e. the noiselike disturbance
 each coordinate of $A^*y$ experiences
 from the \emph{presence} of all the other `weakly interacting'
 coordinates. The thresholding iteration suppresses
this interference in the sparse case by detecting the many `silent'
channels and setting them a priori to zero,
producing a putatively better guess at the next iteration.
At that iteration, the remaining interference is proportional
not to the size of the estimand, but instead to the estimation
error, i.e. it is caused by the
{\it errors} in reconstructing all the weakly interacting coordinates;
these errors are only a fraction of the sizes of the estimands
and so the error is significantly reduced at the next iteration.
%
%
\subsection{ State Evolution}

The above `sparse denoising'/`interference suppression' heuristic,
does agree qualitatively with the
actual behavior one can observe in sample
reconstructions. It is very tempting to
take it literally.
Assuming it is literally true that the MAI is Gaussian
and independent from iteration to iteration,
we can can formally track the evolution, from
iteration to iteration, of the mean-squared error.
\begin{figure}[t]
\centerline{\includegraphics[width=1\linewidth]{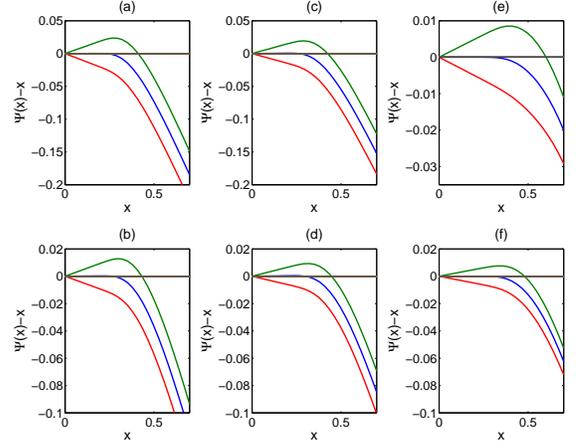}}
\caption{Development of fixed points for formal MSE evolution.
Here we plot $\Psi(\sigma^2)-\sigma^2$ where $\Psi(\,\cdot\,)$
is the MSE map for
$\pb = \sT$ (left column), $\pb=\sC$ (center column) and $\pb=\sQ$
(right column), $\delta = 0.1$ (upper row,$\pb \in \{\sT , \sC\}$), $\delta=0.55$ (upper row,$\pb ={\sQ}$), $\delta = 0.4$
(lower row,$\pb \in \{\sT , \sC\}$) and $\delta=0.75$ (lower row,$\pb =\sQ $). A crossing of the y-axis corresponds to a fixed point of $\Psi$.
If the graphed quantity is negative for positive $\sigma^2$,
$\Psi$ has no fixed points for $\sigma > 0$.
Different curves correspond to different
values of $\rho$: where  $\rho$ is respectively less than,
equal to and greater than $\rho_{\SE}$. In each case, $\Psi$ has a stable fixed
fixed point at zero for $\rho < \rho_{\SE}$, and no other fixed points,
an unstable fixed point at zero for $\rho = \rho_{\SE}$ and
devlops two fixed points at $\rho > \rho_{\SE}$.
Blue curves correspond to $\rho = \rho_{\SE}(\delta;\pb)$,
green to $\rho = 1.05\cdot\rho_{\SE}(\delta;\pb)$,
red to $\rho = 0.95\cdot\rho_{\SE}(\delta;\pb)$.}
\end{figure}

This gives a recursive equation for the {\it formal MSE},
i.e. the MSE which would be true if the heuristic were
true. This takes the form
\begin{eqnarray}
\sigma_{t+1}^2 & = & \Psi(\sigma_t^2)\, ,\label{eq:MSEmap}\\
\Psi(\sigma^2) & \equiv & \E\Big\{\big[\eta\big(X+\frac{\sigma}{\sqrt{\delta}}Z;
\lambda \sigma\big)-X\big]^2\Big\}\, .
\end{eqnarray}
Here expectation is with respect to independent random variables
$Z \sim  \sN(0,1)$ and $X$, whose distribution coincides with the empirical
distribution of the entries of $x_0$.
We use soft thresholding  (\ref{eq:SoftThresh})
if the signal is sparse and signed, i.e. if $\pb=\sC$.
In the case of sparse non-negative vectors, $\pb=\sT$, we will
let $\eta(u;\lambda\sigma,\sT) = \max(u-\lambda\sigma,0)$.
Finally, for $\pb=\sQ$, we let $\eta(u;\sQ) = \sign(u)\, \min(|u|,1)$.
Calculations of this sort are familiar
from the theory of
soft thresholding of sparse signals; see the Supplement
for details.

We call
$\Psi:\sigma^2\mapsto \Psi(\sigma^2)$ the {\it MSE map}.
\begin{definition}
Given implicit parameters  $(\pb,\delta,\rho,\lambda, F)$, with
$F= F_X$  the distribution of the random variable $X$.
\emph{State Evolution}  is the recursive map
(one-dimensional dynamical system):
$\sigma^2_t \mapsto \Psi(\sigma^2_t)$.

Implicit parameters $(\pb,\delta,\rho,\lambda, F)$
stay fixed during the evolution. Equivalently, the full state
evolves by the rule
\[
(\sigma^2_t; \pb, \delta, \rho, \lambda,F_X) \mapsto (\Psi(\sigma^2_t); \pb, \delta, \rho, \lambda,F_X)
\, .
\]
\end{definition}

Parameter space is partitioned into two regions:

\vspace{0.14cm}

\noindent \emph{Region (I)}:  $\Psi(\sigma^2)<\sigma^2$ for all $\sigma^2 \in (0, \E X^2]$.
Here $\sigma_t^2\to 0$
as $t\to \infty$: the SE converges to zero.

\vspace{0.08cm}

\noindent \emph{Region (II)}:  The complement of Region (I). Here,
 the SE recursion does {\em not} evolve to $\sigma^2=0$.

\vspace{0.14cm}

The partitioning of parameter space induces a notion of sparsity threshold,
the minimal sparsity guarantee needed to obtain convergence
of the formal MSE:
\begin{eqnarray}
\rho_{\SE}(\delta;\pb,\lambda,F_X) \equiv
\sup\left\{\rho\, :\;  (\delta,\rho,\lambda,F_X)
\in \mbox{Region (I)} \right\} \, .
\end{eqnarray}
The subscript $\SE$ stands for State Evolution.
Of course, $\rho_{\SE}$ depends on the
case $\pb \in \{\sT,\sC,\sQ\}$; it also
 seems to depend also on the signal distribution $F_X$;
however, an essential simplification is provided by
\begin{proposition}\label{propo:Independence}
For the three canonical problems $\pb\in \{\sT,\sC,\sQ\}$,
any $\delta\in[0,1]$, and any random variable $X$ with
the prescribed sparsity and bounded second moment,
$\rho_{\SE}(\delta;\pb,\lambda,F_X)$ is independent of $F_X$.
\end{proposition}
Independence from $F$ allows us to
write $\rho_{\SE}(\delta;\pb,\lambda)$
for the sparsity thresholds. The proof of this statement
is sketched below, along with the derivation of
a more explicit expression. Adopt the notation
\begin{eqnarray}
\rho_{\SE}(\delta;\pb) = \sup_{\lambda\ge 0}\rho_{\SE}(\delta;\pb,\lambda).
\label{eq:OptimalTune}
\end{eqnarray}
High precision numerical evaluations of such expression uncovers the
following very suggestive
\begin{finding}\label{Finding:IterativePT}
For the three canonical problems $\pb\in \{\sT,\sC,\sQ\}$,
and for any $\delta\in (0,1)$
\begin{eqnarray}
\rho_{\SE}(\delta;\pb) = \rho_{\CG}(\delta;\pb)\, .
\label{eq:IterPolytope}
\end{eqnarray}
\end{finding}

In short, the formal MSE evolves to zero exactly over the
same region of $(\delta,\rho)$ phase space as does
the phase diagram for the corresponding convex optimization!
%
%
\subsection{Failure of standard iterative algorithms}

If we trusted that formal MSE truly describes
the evolution of the iterative thresholding algorithm,
Finding \ref{Finding:IterativePT} would imply that
iterative thresholding allows to undersample
just as aggressively in solving underdetermined
linear systems as the corresponding LP.

Finding \ref{Finding:IterativePT}
gives new reason to hope for a possibility that has already inspired
many researchers over the last five years: the possibility of
finding a very fast algorithm that replicates the behavior
of convex optimization in settings $\sT,\sC,\sQ$.

Unhappily the formal MSE calculation
does not describe the behavior of iterative thresholding:

\vspace{0.14cm}

\noindent \emph{1.} State Evolution does not predict the observed
properties of iterative thresholding algorithms.

\vspace{0.08cm}

\noindent \emph{2.} Iterative thresholding algorithms,
even when optimally tuned, do not achieve
the optimal phase diagram.

\vspace{0.14cm}

In \cite{MaDo09sp}, two of the authors carried out
an extensive empirical study of iterative thresholding algorithms.
Even optimizing over the free parameter $\lambda$ and the nonlinearity $\eta$
the phase transition was observed at significantly smaller values of
$\rho$ than those observed for LP-based algorithms.

Numerical simulations also show very clearly that the
MSE map \emph{does not} describe the evolution of the
actual MSE under iterative thresholding.
The mathematical reason for this failure is quite simple.
After the first iteration, the entries of $x^t$ become strongly dependent,
and State Evolution does not
predict the moments of $x^t$.
%
%
\subsection{Message Passing Algorithm}

The main surprise of this paper is that this failure is not
the end of the story. We now consider a
modification of iterative thresholding inspired by
message passing algorithms for inference in graphical models
\cite{Pearl},
and graph-based error correcting codes \cite{GallagerThesis,RiUBook}.
These are iterative algorithms, whose basic variables (`messages')
are associated to directed edges in a graph that encodes the structure
of the statistical model. The relevant graph here
is a complete bipartite graph over $N$ nodes on one side
(`variable nodes'), and $n$ on the others (`measurement nodes').
Messages are updated according to the rules
\begin{eqnarray}
x^{t+1}_{i\to a} & = &\eta_t\Big(\sum_{b\in [n]\setminus a}A_{bi}z^t_{b\to i}
\Big)\, ,\label{eq:MP1}
\end{eqnarray}
\begin{eqnarray}
z_{a\to i}^t & = &y_a-\sum_{j\in [p]\setminus i} A_{aj}x^t_{j\to a}\, ,
\label{eq:MP2}
\end{eqnarray}
for each $(i,a)\in [N]\times [n]$. We will refer to this
algorithm\footnote{For earlier applications of MP to compressed sensing
see \cite{BraidPractice,Baraniuk,Pfister}. Relations between MP
and LP were explored in a number of papers, see for instance
\cite{Wainwright05,Bayati08}, albeit from a different perspective.} as to
MP.

MP has one important drawback with respect to iterative thresholding.
Instead of updating $N$ estimates, at each iterations we need
to update $Nn$ messages, thus increasing significantly the algorithm
complexity.
On the other hand, it is easy to see that the right-hand side of
eqn~[\ref{eq:MP1}] depends weakly on the index $a$
(only one out of $n$ terms is excluded) and that the right-hand side
of eqn~[\ref{eq:MP1}] depends weakly on $i$. Neglecting
altogether this dependence leads to the iterative
thresholding equations [\ref{eq:IT1}], [\ref{eq:IT2}].
A more careful analysis of this dependence
leads to corrections of order one in the high-dimensional limit.
Such corrections are however fully captured by the last term on the
right hand side of eqn [\ref{eq:FOAMP2}], thus leading to the AMP
algorithm. Statistical physicists would call this
 the `Onsager reaction term'; see \cite{TAP}.
\begin{figure}[t]
\centerline{\includegraphics[width=1.05\linewidth]{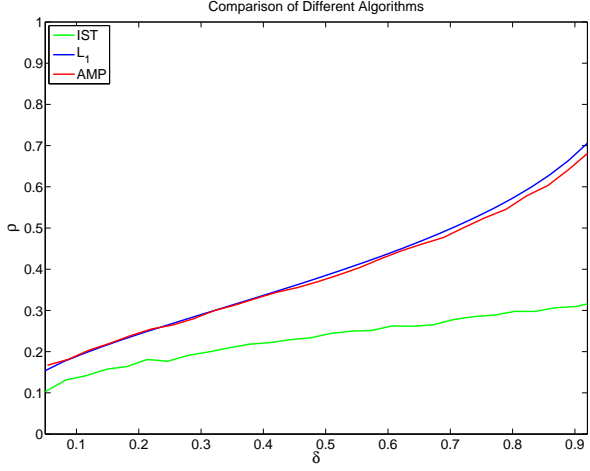}}
\caption{Observed phase transitions of reconstruction algorithms.
Algorithms studied include iterative soft and hard thresholding,
orthogonal matching pursuit, and related.
Parameters of each algorithm are tuned to achieve the best possible phase transition \cite{MaDo09sp}.
Reconstructions signal length $N=1000$. Iterative thresholding algorithms used $T=1000$ iterations. Phase transition curve displays the value of $\rho=k/n$ at which
success rate is 50\%.}
\end{figure}

%
%
\subsection{State Evolution is Correct for MP}

Although AMP seems very similar to simple iterative
thresholding [\ref{eq:IT1}]-[\ref{eq:IT2}], SE
accurately describes its properties, but not those of
the standard iteration.
As a consequence of Finding \ref{Finding:IterativePT},
properly tuned versions of MP-based algorithms
are asymptotically as powerful as
LP reconstruction.

We have conducted
extensive simulation experiments with AMP,
and more limited experiments with MP, which is
computationally more intensive (for details see the complementary material).
These experiments show  that  the performance
of the algorithms can be accurately
modeled using the MSE map. Let's be more specific.

According to SE, performance of the AMP algorithm is predicted by
tracking the evolution of the formal MSE $\sigma^2_t$
  via the recursion [\ref{eq:MSEmap}]. Although this formalism is
quite simple, it is accurate in the high dimensional
limit. Corresponding to the formal quantities calculated
by SE are the actual quantities, so of course
to the formal MSE corresponds the true MSE $N^{-1} \| x^t - x_0 \|_2^2$.
 Other quantities can be computed in terms
of the state $\sigma^2_t$ as well: for instance
the true false alarm rate  $(N-k)^{-1} \# \{ i : x^t(i)
\neq 0 \; \mbox{ and } \; x_0(i) = 0 \} $ is predicted via the formal false
alarm rate $\prob\{ \eta_t(X + \delta^{-1/2} \sigma_t Z ) \neq 0 | X = 0 \}$.
Analogously,
the true missed-detection rate  $k^{-1} \# \{ i : x^t(i) = 0 \;
\mbox{ and } \;
x_0(i) \neq 0 \} $ is predicted by
the formal missed-detection rate
$\prob\{ \eta_t(X + \delta^{-1/2} \sigma_t Z ) =  0 | X \neq 0 \}$, and so on.

Our experiments establish agreement of
actual and formal quantities.
\begin{finding}  For the AMP algorithm, and large dimensions
$N,n$, we observe

\vspace{0.1cm}

\noindent\emph{I.}  SE
correctly predicts the evolution
of numerous statistical properties of $x^t$  with the iteration number
$t$.  The MSE, the number of
nonzeros in $x^t$, the number of false alarms,
the number of missed detections,  and several other
measures all evolve in way that matches the
state evolution formalism to within experimental accuracy.

\vspace{0.1cm}

\noindent\emph{II.} SE correctly predicts the success/failure
to converge to the correct result. In particular, SE predicts no convergence
when $\rho > \rho_{\SE}(\delta;\pb,\lambda)$, and convergence if
$\rho < \rho_{\SE}(\delta;\pb,\lambda)$.
This is indeed observed
empirically.
\end{finding}
Analogous observations were made for MP.

\subsection{Optimizing the MP Phase Transition}\,

An inappropriately tuned version of MP/AMP
will not perform well compared to other algorithms,
for example LP-based reconstructions.
However, SE provides a natural strategy to tune MP and
AMP (i.e. to choose the free parameter $\lambda$):
simply use the value achieving the maximum in eqn~[\ref{eq:OptimalTune}].
We denote this value by $\lambda_\pb(\delta)$, $\pb\in\{\sT,\sC,\sQ\}$,
and refer to the resulting algorithms as to
\emph{optimally tuned MP/AMP} (or sometimes MP/AMP for short).
They achieve the State Evolution phase transition:
\[
   \rho_{\SE}(\delta;\pb) = \rho_{\SE}(\delta; \pb, \lambda_\pb(\delta)) .
\]
An explicit characterization of $\lambda_\pb(\delta)$,
$\pb\in\{\sT, \sC\}$ can be found in the next section.

We summarize below the properties of optimally tuned AMP/MP
within the SE formalism.
\begin{theorem}\label{Claim:FOAMP_PT}
For $\delta\in[0,1]$, $\rho < \rho_{\SE}(\delta;\pb)$, and
any associated random variable $X$,
the formal MSE of optimally-tuned AMP/MP evolves to zero under SE.
Viceversa, if  $\rho>\rho_{\SE}(\delta;\pb)$, the formal MSE
does not evolve to zero.
Further, for  $\rho< \rho_{\SE}(\delta;\pb)$, there exists
$b = b(\delta, \rho)>0$ with the following property.
If $\sigma_t^2$ denotes the formal MSE after $t$ SE steps, then, for all $t\ge 0$
\begin{equation} \label{eq:CvgExpon}
\sigma_t^2\le \sigma_0^2\exp(-bt)  .
\end{equation}
\end{theorem}

%
%
%
%
\section{Details About the MSE Mapping}

In this section, we sketch the proof of Proposition
\ref{propo:Independence}: the iterative threshold does not depend
on the details of the signal distribution. Further, we show how
to derive the explicit expression for $\rho_{\SE}(\delta;\pb)$,
$\pb\in\{\sT,\sC\}$, given in the introduction.
%
%
\subsection{Local Stability Bound}

The state evolution threshold $\rho_{\SE}(\delta;\pb,\lambda)$ is the
supremum of all $\rho$'s such that the MSE map
$\Psi(\sigma^2)$ lies below the $\sigma^2$ line for all
$\sigma^2>0$. Since $\Psi(0) = 0$, for this
to happen it must be true that the derivative of the MSE map at
$\sigma^2=0$ smaller than or equal to $1$.
We are therefore led to define the following
`local stability' threshold:
\begin{eqnarray}
\rho_{\LS}(\delta;\pb,\lambda)
\equiv  \sup\left\{\rho\, :\; \left.\frac{\de \Psi}{\de \sigma^2}
\right|_{\sigma^2=0}<1 \right\}\, .
\end{eqnarray}
The above argument implies that
$\rho_{\SE}(\delta;\pb,\lambda)\le\rho_{\LS}(\delta;\pb,\lambda)$.

Considering for instance $\pb=\sT$, we obtain the following
expression for the first derivative of $\Psi$
\begin{eqnarray*}
\frac{\de \Psi}{\de \sigma^2} = \left(\frac{1}{\delta}+
\lambda^2\right)\E\, \Phi\Big(\frac{\sqrt{\delta}}
{\sigma}(X-\lambda\sigma)\Big)-\frac{\lambda}{\sqrt{\delta}}
\E\, \phi\Big(\frac{\sqrt{\delta}}{\sigma}(X-\lambda\sigma)\Big)\, ,
\end{eqnarray*}
where $\phi(z)$ is the standard Gaussian density at $z$
and $\Phi(z) = \int_{-\infty}^z\phi(z')\,\de z'$ is the Gaussian distribution.

Evaluating this expression as $\sigma^2\downarrow 0$, we get the local
stability threshold for $\pb=\sT$:
\begin{eqnarray*}
\rho_{\LS}(\delta;\pb,\lambda)  =  \left.
\frac{1-(\kappa_\pb/\delta)\big[(1+z^2)\Phi(-z)-z\phi(z)\big]}{
1+z^2-\kappa_{\pb}
\big[(1+z^2)\Phi(-z)-z\phi(z)\big]}\right|_{z = \lambda\sqrt{\delta}}\, ,
\end{eqnarray*}
where $\kappa_{\pb}$ is the same as in [\ref{eq:rhose}].
Notice that $\rho_{\LS}(\delta;\sT,\lambda)$ depends
on the distribution of $X$ only through its sparsity
(i.e. it is independent of $F_X$).
%
%
\subsection{Tightness of the Bound and Optimal Tuning}

We argued that $\left.\frac{\de \Psi}{\de \sigma^2}
\right|_{\sigma^2=0}<1$ is necessary
for the MSE map to converge to $0$.
This condition turns out to be sufficient because
the function $\sigma^2\mapsto \Psi(\sigma^2)$ is concave on
$\reals_+$. This indeed yields
\begin{eqnarray}
\sigma_{t+1}^2\le \left.\frac{\de \Psi}{\de \sigma^2}
\right|_{\sigma^2=0}\, \sigma_t^2\, ,
\end{eqnarray}
which implies exponential convergence to the correct solution
[\ref{eq:CvgExpon}]. In particular we have
\begin{eqnarray}
\rho_{\SE}(\delta;\pb,\lambda) = \rho_{\LS}(\delta;\pb,\lambda)\, ,
\end{eqnarray}
whence $\rho_{\SE}(\delta;\pb,\lambda)$ is independent of $F_X$ as claimed.

To prove $\sigma^2\mapsto \Psi(\sigma^2)$ is concave, one proceeds by
computing its second derivative. For instance, in the
case $\pb=\sT$, one needs to differentiate the expression given above
for the first derivative. We omit details but point out two useful remark:
$(i)$ The contribution due to $X=0$ vanishes; $(ii)$
Since a convex combination of concave functions is also concave,
it is sufficient to consider the case in which $X=x_*$ deterministically.

As a byproduct of this argument we obtain explicit expressions for
the optimal tuning parameter, by maximizing the local stability threshold
\begin{eqnarray*}
\lambda_{\sT}(\delta) = \frac{1}{\sqrt{\delta}}
\arg\max_{z\ge 0}\left\{
\frac{1-(\kappa_{\pb}/\delta)\big[(1+z^2)\Phi(-z)-z\phi(z)\big]}{
1+z^2-\kappa_{\pb}\big[(1+z^2)\Phi(-z)-z\phi(z)\big]}\right\}\, .
\end{eqnarray*}
Before applying this formula in practice, please read the important notice
in Supplemental Information.
%
\section{Discussion}

\subsection{Relation with Minimax Risk}

Let $\cF^\pm_{\eps}$ denote the class of
probability distributions $F$ supported on $(-\infty,\infty)$
with $\prob\{X\neq 0\}\le \eps$,
and let $\eta(x;\lambda,\sC)$
denote the soft-threshold function [\ref{eq:SoftThresh}] with
threshold value $\lambda$. The minimax risk \cite{DJ94a} is defined as
\begin{eqnarray}
M^\pm(\eps) \equiv \inf_{\lambda\ge 0} \sup_{F\in \cF^{\pm}_{\eps}}
\E_F\{[\eta(X+Z;\lambda,\sC)-X]^2\}\, ,
\end{eqnarray}
with $\lambda^\pm(\eps)$ the optimal $\lambda$.
The optimal SE phase transition
and optimal SE threshold obey
\beq \label{eq:connect}
\delta = M^\pm(\rho\delta)\, ,
\;\;\;\;\;\;\rho = \rho_{\SE}(\delta;\sC).
\eeq
An analogous relation holds between the positive case $\rho_{\SE}(\delta;\sT)$,
and the minimax threshold risk $M^+$ where $F$ is constrained to be
a distribution on $[0,\infty)$.  Exploiting [\ref{eq:connect}],
Supporting Information proves that
\[
   \rho_{\CG}(\delta) = \rho_{\SE}(\delta)( 1+ o(1)) , \qquad \delta \goto 0.
\]

\subsection{Other Message Passing Algorithms}

The nonlinearity $\eta(\,\cdot\,)$ in AMP
eqns [\ref{eq:FOAMP1}], [\ref{eq:FOAMP2}]
might be chosen differently.  For sufficiently regular
such choices, the SE formalism might predict evolution of the MSE.
One might hope to use SE to design `better' threshold nonlinearities.

%
The threshold functions used here are such that
the MSE map $\sigma^2\mapsto \Psi(\sigma^2)$
is monotone and concave.
As a consequence, the phase transition line  $\rho_{\SE}(\delta;{\pb})$
for optimally tuned AMP is independent of the empirical distribution of
the vector $x_0$. State Evolution may be inaccurate without such
properties.

Where SE is accurate, it offers limited room for improvement
over the results here. If $\tilde{\rho}_{\SE}$ denotes a (hypothetical)
phase transition derived by SE with {\it any nonlinearity} whatsoever,
Supporting Information exploits [\ref{eq:connect}] to prove
\[
   \tilde{\rho}_{\SE}(\delta;\pb) \leq \rho_{\SE}(\delta;\pb)( 1+ o(1)) , \qquad \delta \goto 0\, , \quad \pb \in \{\sT,\sC\}\, .
\]
In the limit of high undersampling, the nonlinearities studied here
offer essentially unimprovable SE phase transitions.
Our reconstruction experiments also suggest that other nonlinearities yield
little improvement over thresholds used here.

%
%
\subsection{Universality}

The SE-derived phase transitions are not sensitive to the
detailed distribution of coefficient amplitudes. Empirical
results in Supporting Information find similar
insensitivity of observed phase transitions for MP.

Gaussianity of the measurement matrix $A$
can be relaxed; Supporting Information
finds that other random matrix ensembles
exhibit comparable phase transitions.

In applications, one often uses very large matrices $A$
which are never explicitly represented, but only applied as
operators; examples include randomly undersampled
partial Fourier transforms.
Supporting Information finds that  observed phase transitions for MP
in the partial Fourier case
are comparable to those for random $A$.

%
%

%
%

%
%
%

\section*{Acknowledgements}
A. Montanari was partially supported by the NSF CAREER award CCF-0743978
and the NSF grant DMS-0806211, and thanks Microsoft Research New England
for hospitality during completion of this work. A. Maleki was partially supported by NSF DMS-050530.

%
%


%
%
\appendix
\subsection{Important Notice}
Readers familiar with the  literature of thresholding
of sparse signals will want to know that an implicit rescaling is needed to match equations
from that literature with equations here.  Specifically, in the traditional literature,
one is used to seeing expressions $\eta( x ; \lambda \sigma )$ in cases where $\sigma$ is the
standard deviation of an underlying normal distribution. This means the threshold $\lambda$
is specified in standard deviations, so many people will immediately understand values like
of $\lambda=2, 3$ etc in terms of their false alarm rates.
In the main text, the expression $\eta( x ; \lambda \sigma )$ appears numerous times,
but note that $\sigma$ is not the standard deviation of the relevant normal distribution;
instead, the standard deviation of that normal is $\tau = \sigma/\sqrt{\delta}$.  It follows
that $\lambda$ in the main text is calibrated differently from the way
$\lambda$ would be calibrated in other sources, differing by a $\delta$-dependent scale factor.
If we let $\lambda_{SE}^{sd}$ denote the quantity $\lambda_{SE}$ appropriately rescaled
so that it is in units of standard deviations of the underlying normal distribution, then
the needed conversion to sd units is
\beq \label{eq:sdunits}
 \lambda_{SE}^{sd} = \lambda_{SE} \cdot \sqrt{\delta} .
\eeq

\subsection{A summary of notation}
The main paper will be referred as DMM throughout this note. All the notations are consistent with the notations used in DMM. We will use repeatedly the notation $\eps=\delta \rho$.

%
%

\subsection{State Evolution Formulas}
In the main text we mentioned $\rho_{\SE}(\delta;\gly, \lambda, F_X)$ is independent of $F_X$. We also mentioned a few formulas for $\rho_{\SE}(\delta;\gly)$. The goal of this section is to explain the calculations involved in deriving these results. First, recall
the expression for the MSE map
\begin{eqnarray}
\Psi( \sigma^2 )  =
\E \Big\{
\big(\eta( X + \frac{\sigma}{\sqrt{\delta}} Z; \lambda\sigma,\gly) - X\big)^2
\Big\} \, .
\end{eqnarray}
We denote by $\partial_1\eta$ and $\partial_2 \eta$ the
partial derivatives of $\eta$ with respect to its
first and second arguments. Using Stein's lemma and the
fact that $\partial_1^2\eta(x;y,\gly) = 0$ almost everywhere, we get
\begin{eqnarray}
\frac{\de \Psi}{\de \sigma^2} = \frac{1}{\delta}\,
\E\Big\{\partial_1 \eta(X+\frac{\sigma}{\sqrt{\delta}}Z;\lambda\sigma)^2
\Big\}+\nonumber \\
\frac{\lambda}{\sigma}\E\Big\{
\big[\eta(X+\frac{\sigma}{\sqrt{\delta}}Z;\lambda\sigma)
-X\big]\partial_2 \eta
(X+\frac{\sigma}{\sqrt{\delta}}Z;\lambda\sigma)
\Big\}\, ,
\end{eqnarray}
where we dropped the dependence of $\eta(\,\cdot\,)$ on the constraint
$\gly$ to simplify the formula.
%
\subsubsection{Case $\gly=+$}

In this case we have $X\ge 0$  almost surely and the threshold function
is
\begin{eqnarray}
\eta(x;\lambda\sigma) = \left\{
\begin{array}{ll}
(x-\lambda\sigma) & \mbox{ if $x\ge\lambda\sigma$,}\\
0 & \mbox{ otherwise.} \nonumber
\end{array}\right.
\end{eqnarray}
As a consequence $\partial_1\eta(x;\lambda\sigma)=-
\partial_2\eta(x;\lambda\sigma) = \ind(x\ge \lambda\sigma)$
(almost everywhere).
This yields
\begin{eqnarray}
\frac{\de \Psi}{\de \sigma^2} = \left(\frac{1}{\delta}+
\lambda^2\right)\E\, \Phi\Big(\frac{\sqrt{\delta}}
{\sigma}(X-\lambda\sigma)\Big)\nonumber \\
-\frac{\lambda}{\sqrt{\delta}}
\E\, \phi\Big(\frac{\sqrt{\delta}}{\sigma}(X-\lambda\sigma)\Big)\, . \nonumber
\end{eqnarray}
As $\sigma\downarrow 0$, we have $\Phi\Big(\frac{\sqrt{\delta}}
{\sigma}(X-\lambda\sigma)\Big)\to 1$
and $\phi\Big(\frac{\sqrt{\delta}}
{\sigma}(X-\lambda\sigma)\Big)\to 0$ if $X>0$. Therefore,
\begin{eqnarray}
\left.\frac{\de \Psi}{\de \sigma^2}\right|_0 =
\left(\frac{1}{\delta}+
\lambda^2\right)\rho\delta+ \left(\frac{1}{\delta}+
\lambda^2\right)\, (1-\rho\delta)\, \Phi(-\lambda\sqrt{\delta})
\nonumber \\
-\frac{\lambda}{\sqrt{\delta}}(1-\rho\delta)\,
\phi(-\lambda\sqrt{\delta})\, . \nonumber
\end{eqnarray}
The local stability threshold $\rho_{\LS}(\delta;+,\lambda)$
is obtained by setting $\left.\frac{\de \Psi}{\de \sigma^2}\right|_0=1$.

In order to prove the concavity of $\sigma^2\mapsto\Psi(\sigma^2)$ first notice that a convex combination
of concave functions is concave and so it is sufficient to show the
concavity in the case $X=x\ge 0$ deterministically. Next
notice that, in the case $x=0$, $\frac{\de \Psi}{\de \sigma^2}$
is independent of $\sigma^2$. A a consequence, it is sufficient
to prove $\frac{\de^2 \Psi_{x}}{\de (\sigma^2)^2}\le 0$
where

\begin{eqnarray}
\delta\, \frac{\de \Psi_{x}}{\de \sigma^2} = \left(1+
\lambda^2\delta\right) \Phi\Big(\frac{\sqrt{\delta}}
{\sigma}(x-\lambda\sigma)\Big)-\lambda\sqrt{\delta}
\; \phi\Big(\frac{\sqrt{\delta}}{\sigma}(x-\lambda\sigma)\Big)\, . \nonumber
\end{eqnarray}

Using $\Phi'(u) = \phi(u)$ and $\phi'(u) = -u\phi(u)$, we get
%
%
\begin{eqnarray}
\delta\,
\frac{\de^2 \Psi_{x}}{\de (\sigma^2)^2} =  -\frac{x}{2\sigma^3}
\left\{1+\frac{\lambda\delta}{\sigma}\,x
\,
\right\}\phi\Big(\frac{\sqrt{\delta}}{\sigma}(x-\lambda\sigma)\Big)< 0
\end{eqnarray}
for $x>0$.
%
%
\subsubsection{Case $\gly=\pm$}

Here $X$ is supported on $(-\infty,\infty)$ with
$\prob\{X\neq 0\}\le\eps=\rho\delta$. Recall the definition of soft threshold
\begin{eqnarray}
\eta(x;\lambda\sigma) = \left\{
\begin{array}{ll}
(x-\lambda\sigma) & \mbox{ if $x\ge\lambda\sigma$,}\\
(x+\lambda\sigma) & \mbox{ if $x\le-\lambda\sigma$,}\\
0 & \mbox{ otherwise.} \nonumber
\end{array}\right.
\end{eqnarray}
As a consequence
$\partial_1\eta(x;\lambda\sigma)=\ind(|x|\ge \lambda\sigma)$ and
$\partial_2\eta(x;\lambda\sigma)=-\sign(x)\ind(|x|\ge \lambda\sigma)$.
This yields
\begin{eqnarray}
\frac{\de \Psi}{\de \sigma^2} & = & \left(\frac{1}{\delta}+
\lambda^2\right)\E\Big\{
\Phi\Big(\frac{\sqrt{\delta}}
{\sigma}(X-\lambda\sigma)\Big)+ \nonumber \\
&&\Phi\Big(-\frac{\sqrt{\delta}}
{\sigma}(X+\lambda\sigma)\Big)\Big\}\nonumber\\
&&-\frac{\lambda}{\sqrt{\delta}}\,
\E\Big\{
\phi\Big(\frac{\sqrt{\delta}}{\sigma}(X-\lambda\sigma)\Big)
+\phi\Big(\frac{\sqrt{\delta}}{\sigma}(X+\lambda\sigma)\Big)\Big\}\, . \nonumber
\end{eqnarray}
By letting $\sigma\downarrow 0$ we get
\begin{eqnarray}
\left.\frac{\de \Psi}{\de \sigma^2}\right|_0 =
\left(\frac{1}{\delta}+
\lambda^2\right)\rho\delta+ \left(\frac{1}{\delta}+
\lambda^2\right)\, (1-\rho\delta)\, 2\, \Phi(-\lambda\sqrt{\delta}) \nonumber \\
-\frac{\lambda}{\sqrt{\delta}}(1-\rho\delta)\,2\,
\phi(-\lambda\sqrt{\delta})\, , \nonumber
\end{eqnarray}
which yields the local stability threshold $\rho_{\LS}(\delta;\pm,\lambda)$
by $\left.\frac{\de \Psi}{\de \sigma^2}\right|_0=1$.

Finally the proof of the concavity of $\sigma^2\mapsto\Psi(\sigma^2)$
is completely analogous to the case $\gly=+$.
%
%
\subsubsection{Case $\gly=\Box$}
Finally consider the case of $X$ supported on $[-1,+1]$ with
$\prob\{X\not\in \{+1,-1\}\}\le\eps$. In this case we proposed the following nonlinearity,
\begin{eqnarray}
\eta(x) = \left\{
\begin{array}{ll}
+1 & \mbox{ if $x> +1$,}\\
x & \mbox{ if $-1\le x\le +1$,}\\
-1 & \mbox{ if $x\le -1$.} \nonumber
\end{array}\right.
\end{eqnarray}
Notice that the nonlinearity does not depend
on any threshold parameter. Since $\partial_1\eta(x) = \ind(x\in [-1,+1])$,
\begin{eqnarray}
\frac{\de \Psi}{\de \sigma^2} &=& \frac{1}{\delta}\,
\prob\Big\{X+\frac{\sigma}{\sqrt{\delta}}Z\in [-1,+1]
\Big\} \nonumber \\ 
& = &\frac{1}{\delta}\,
\E\Big\{\Phi\Big(\frac{\sqrt{\delta}}{\sigma}(1-X)\Big)
-\Phi\Big(-\frac{\sqrt{\delta}}{\sigma}(1+X)\Big)\Big\}\, . \nonumber
\end{eqnarray}
As $\sigma\downarrow 0$ we get
\begin{eqnarray}
\left.\frac{\de \Psi}{\de \sigma^2}\right|_0 = \frac{1}{2\delta}
(1+\rho\delta)\, , \nonumber
\end{eqnarray}
whence the local stability condition $\left.\frac{\de \Psi}{\de \sigma^2}
\right|_0< 1$ yields $\rho_{\LS}(\delta;\Box) = (2-\delta^{-1})_+$.

Concavity of $\sigma^2\mapsto \Psi(\sigma^2)$ immediately follows from
the fact that $\Phi(\frac{\sqrt{\delta}}{\sigma}(1-x))$
is non-increasing in $\sigma$ for $x\le 1$ and
$\Phi(-\frac{\sqrt{\delta}}{\sigma}(1+x))$ is non-decreasing for $x\ge -1$.
Using the combinatorial geometry result of \cite{DoTa08ArXiv} we get
\begin{theorem}
For any $\delta\in[0,1]$,
\begin{eqnarray}
\rho_{\CG}(\delta;\Box) =\rho_{\SE}(\delta;\Box) =\rho_{\LS}(\delta;\Box) =
\max\big\{ 0 , 2-\delta^{-1}\big\}\,.
\end{eqnarray}
\end{theorem}
%
%
\subsection{Relation to Minimax Thresholding}

\subsubsection{Minimax Thresholding Policy}

We denote by $\calF_\eps^+$ the collection of all CDF's supported in
$[0,\infty)$ and with $F(0) \geq 1-\eps$, and by $\calF_\eps^\pm$
the collection of all CDF's supported in $(-\infty,\infty)$
and with $F(0+)-F(0-) \geq 1-\eps$.
For $\gly \in \{ +, \pm \}$, define the minimax threshold MSE
\begin{eqnarray}
M^*(\eps;\gly) = \inf_\lambda \sup_{F\in \calF^\gly_\eps}
\E_F \left\{ \eta( X + Z; \lambda,\gly) - X)^2 \right\}  \, ,\label{eq:MinmaxDef}
\end{eqnarray}
where $\E_F$ denote expectation with respect to the random variable
$X$ with distribution $F$, and $\eta(x;\lambda) = \sign(x)(|x|-\lambda)_+$
for $\gly=\pm$ and $\eta(x;\lambda) = (x-\lambda)_+$ for $\gly = +$.
{\it Minimax Thresholding} was discussed  for the case
$\gly = +$ in \cite{DJHS92} and  for $\gly=\pm$ in \cite{DJ94a,DJ94b}.

This machinery gives us a way to look at the
results derived above in very commonsense terms.
Suppose we know $\delta$ and $\rho$ but {\it not} the
distribution $F$ of $X$.  Let's consider what threshold
one might use, and ask at each given iteration
of SE, the threshold which gives us the best possible control
of the resulting formal MSE. That best possible threshold $\lambda^t$
is by definition the minimax threshold at nonzero
fraction $\eps = \rho\delta$, appropriately scaled by the
effective noise level $\tau = \sigma/\sqrt{\delta}$,
\[
        \lambda^t = \lambda^*(\rho \cdot \delta; \gly) \cdot \sigma /\sqrt{\delta} ,
\]
where $ \gly \in \{ + , \pm \} $ depending on the case at hand.
Note that this threshold does not depend on $F$. It depends on iteration
only through the effective noise level at that iteration.
The guarantee we then get for the formal MSE is
the minimax threshold risk, appropriately scaled by the
square of the effective noise level:
\begin{eqnarray}
      \MSE \leq M^*(\rho\delta; \gly) \cdot \tau^2 = M^*(\rho\delta; \gly)
\frac{\sigma^2}{\delta} ,\; .\label{eq:LastMSE}
\end{eqnarray}
for  $\gly \in \{ + , \pm \}$.
This guarantee gives us a reduction in MSE over the previous iteration
if and only if the right-hand side in Eq.~(\ref{eq:LastMSE})
is smaller than $\sigma^2$, i.e. if and only if
\[
    M^*(\rho\delta; \gly) < \delta \, ,\;\;\;\;\;\; \gly \in \{ + , \pm \} .
\]

In short, we can use state evolution
with the minimax threshold, appropriately scaled
by effective noise level, and we get a guaranteed fractional reduction in MSE
at each iteration, with fractional improvement
\beq \label{eq:mmDynamics-1}
             \omega_{\MM}(\delta,\rho; \gly ) = ( 1 - M^*(\rho\delta; \gly)/\delta ) ;
\eeq
hence the formal SE evolution is bounded by:
\beq \label{eq:mmDynamics-2}
         \sigma_t^2 \leq \omega_{\MM}(\delta,\rho; \gly )^t   \cdot \expect X^2 , \qquad  t=1,2, \dots .
\eeq
Results analogous to those
of the main text hold for this minimax thresholding policy.
That is, we can define a minimax thresholding phase transition
such that below that transition,
state evolution with minimax thresholding
converges:
\[
    \rho_{\MM}(\delta ; \gly) =  \sup\{ \rho: M^*(\rho \delta; \gly) < \delta \} ; \qquad \gly \in \{ +, \pm \} .
\]
\begin{theorem}
Under SE with the minimax thresholding policy described above,
for each $(\delta,\rho)$ in $(0,1)^2$ obeying $\rho < \rho_{\MM}(\delta;\gly)$,
and for every marginal distribution $F \in \cF_\eps^\gly$,
the formal MSE evolves to zero, with dynamics bounded by (\ref{eq:mmDynamics-1})-
(\ref{eq:mmDynamics-2}).
\end{theorem}

\subsubsection{Relating Optimal Thresholding to Minimax Thresholding}
An important difference between the optimal threshold
defined in the main text and the minimax threshold
is that $\lambda_\gly = \lambda_\gly(\delta)$ depends only
on the assumed $\delta$ -- no specific $\rho$ need be chosen
while minimax thresholding as defined above requires
that one specify both $\delta$ and $\rho$.  However,
since the methodology is seemingly pointless above
the minimax phase transition, one might think to
specify $\rho = \rho_{\MM}(\delta; \gly)$. This
new threshold $\lambda_{\MM}(\delta; \gly) = \lambda^*(\delta \rho_{\MM}(\delta); \gly)$
then requires no specification of $\rho$.
As it turns out, the SE threshold coincides
with this new threshold.

\begin{theorem} \label{thm:rhhose-rhomm}
For $\gly\in\{+,\pm\}$ and $\delta\in[0,1]$
\beq \label{eq:rhose-altdef}
    M^*(\rho\delta;\gly) = \delta\;\;\; \;\;\mbox{\sf if and only if }\;\;\;
\rho = \rho_{\SE}(\delta;\gly)\, .
\eeq
Let $\lambda_\gly(\delta)$ denote the minimax threshold defined in the main text, and let $\lambda^{sd}_{\gly}(\delta)$ denote denote the same
quantity expressed in sd units (\ref{eq:sdunits}).
Then
\[
     \lambda^{sd}_\gly(\delta) = \lambda^\gly(\rho\delta), \qquad
      \rho = \rho_{\SE}(\delta;\gly), \qquad \gly \in \{ +, \pm \}
\]
\end{theorem}
\begin{proof}
It is convenient to introduce the following explicit notation for
the MSE map:
\begin{eqnarray}
\Psi( \sigma^2 ; \delta,  \lambda, F)  =
\E_F \Big\{
\big(\eta( X + \frac{\sigma}{\sqrt{\delta}} Z; \lambda\sigma) - X\big)^2
\Big\} \, ,
\end{eqnarray}
where $Z\sim {\sf N}(0,1)$ is independent of $X$, and $X \sim F$.
As above, we drop the dependency of the threshold function on
$\gly\in\{+,\pm\}$
Since $\eta(ax;a\lambda) = a\, \eta(x;\lambda)$ for any positive $a$,
we have the scale invariance
\begin{eqnarray}
\Psi( \sigma^2 ; \delta,  \lambda, F,\gly)  = \frac{\sigma^2}{\delta} \Psi( 1 ; 1,  \lambda\sqrt{\delta},  S_{\delta^{1/2}/\sigma} F ),
\end{eqnarray}
where $(S_a F)(x) = F(x/a)$ is the operator that takes the CDF of an random
variable $X$ and returns the CDF of the random variable $aX$.

Define
\begin{eqnarray}
J(\delta,\rho;\gly) = \inf_{\lambda\ge 0}\sup_{F\in\cF^\gly_{\eps}}
\sup_{\sigma^2\in (0,\E_F\{X^2\}]}\, \frac{1}{\sigma^2}\,
\Psi( \sigma^2 ; \delta,  \lambda, F,\gly)\, ,
\end{eqnarray}
where $\eps\equiv\rho\delta$. It follows from the definition of $\SE$
threshold that $\rho< \rho_{\SE}(\delta;\gly)$ if and only if
$J(\delta,\rho;\gly)< 1$. We first notice that by concavity
of $\sigma^2\mapsto \Psi( \sigma^2 ; \delta,  \lambda, F,\gly)$, we have
\begin{eqnarray}
J(\delta,\rho;\gly) & =& \inf_{\lambda}\sup_{F\in\cF^\gly_{\eps}}
\sup_{\sigma^2 >0}\, \frac{1}{\sigma^2}\,
\Psi( \sigma^2 ; \delta,  \lambda, F,\gly)\\
& =& \frac{1}{\delta}\inf_{\lambda}\sup_{F\in\cF^\gly_{\eps}}
\sup_{\sigma^2 >0}\, \Psi( 1 ; 1,  \lambda\sqrt{\delta},  S_{\delta^{1/2}/\sigma} F )\\
& = & \frac{1}{\delta}\inf_{\lambda}\sup_{F\in\cF^P_{\eps}}\,
\Psi( 1 ; 1,  \lambda,   F )
\end{eqnarray}
where the second identity follows from the invariance property
and the third from the observation that $S_a\cF^\gly_{\eps} = \cF^\gly_{\eps}$
for any $a>0$. Comparing with the definition (\ref{eq:MinmaxDef}),
we finally obtain
\begin{eqnarray}
J(\delta,\rho;\gly) = \frac{1}{\delta}\, M^*(\delta\rho ;\gly)\, .
\end{eqnarray}
Therefore $\rho<\rho_{\SE}(\delta;\gly)$ if and only $\delta>M^*(\delta\rho;\gly)$,
which implies the thesis.
\end{proof}

\subsection{Convergence Rate of State Evolution}

The optimal thresholding policy described in  the main text is the same
as using the minimax thresholding policy
but instead assuming the most pessimistic possible choice
of $\rho$ -- the largest $\rho$ that can possibly
make sense.  In contrast minimax thresholding is $\rho$-adaptive,
and can use a smaller threshold where it would be valuable.
Below the SE phase transition, both methods
will converge, so what's different?

Note that $\lambda_{\SE}(\delta;\gly)$ and $\lambda_{MM}(\delta,\rho;\gly)$ are dimensionally different; $\lambda_{MM}$ is in standard deviation units. Converting $ \lambda_{\SE}$ into  sd units by (\ref{eq:sdunits}), we have
${\lambda}^{sd}_{\SE} = \lambda_{\SE}\cdot \delta^{1/2} $.  Even after this calibration,
we find that  methods will generally use different thresholds, i.e.
if $\rho < \rho_{\SE}$,
\[
    \lambda_{\MM}(\delta,\rho; \gly) \neq  \lambda^{sd}_{\SE}(\delta; \gly) , \qquad \gly \in \{ +,\pm \}.
\]
In consequence, the methods may have different rates of convergence.
Define the worst-case threshold MSE
\[
\MSE(\eps, \lambda;\gly) =  \sup_{F\in \calF^\gly_\eps}
\E_F \left\{ \eta( X + Z; \lambda) - X)^2 \right\}
\]
and set
\[
M_{\SE}(\delta,\rho; \gly ) =  \MSE(\delta\rho, \lambda^{sd}_{\SE}(\delta,\gly) ; \gly).
\]
This is the MSE guarantee achieved by using  $\lambda^{sd}_{\SE}(\delta)$
when in fact $(\delta,\rho)$ is the case.
Now by definition of minimax threshold MSE,
\beq \label{eq:cmpMSE}
    M_{\SE}(\delta,\rho; \gly )  \geq M^*(\delta\rho; \gly );
\eeq
the inequality is generally strict.  The convergence rate
of optimal AMP under SE was described implicitly in the main text.
We can give more precise information using this notation.  Define
\[
\omega_{\SE}(\delta, \rho; \gly) = ( 1- M_{\SE}(\delta,\rho; \gly ) /\delta) ;
\]
Then we have for the formal MSE of AMP
\[
    \sigma^2_t \leq \omega_{\SE}(\delta, \rho; \gly)^t \cdot \E X^2, \qquad t=1,2,3, \dots
\]
In the main text, the same relation was written in terms of $\exp(-bt)$,
with $b > 0$; here we see that  we may take $b(\delta,\rho) = -\log(\omega_{\SE}(\delta,\rho))$.
Explicit evaluation of this $b$ requires evaluation of the worst-case
thersholding risk $\MSE(\eps,\lambda)$.
Now by (\ref{eq:cmpMSE}) we have
\[
  \omega_{\SE}(\delta, \rho; \gly)  \geq  \omega_{\MM}(\delta, \rho; \gly),
\]
generally with strict inequality;
so by using the $\rho$-adaptive threshold one gets better
speed guarantees.
%
%

\subsection{Rigorous Asymptotic Agreement  of SE and CG}
In this section we prove
\begin{theorem} \label{thm:CGSEequiv}
For $\gly \in \{ +,\pm \}$
\begin{eqnarray}
\lim_{\delta \goto 0}\frac{\rho_{\CG}(\delta;\gly)}{\rho_{\SE}(\delta;\gly)}   =  1 .
\end{eqnarray}
\end{theorem}
In words,   $\rho_{\CG}(\delta; \gly)$ is the phase transition computed
by combinatorial geometry (polytope theory) and
$\rho_{\SE}(\delta,\gly)$ obtained by state evolution:
they are rigorously equivalent in the highly undersampled
limit (i.e.  $\delta \goto 0$ limit).  In the main text,
we only can make the observation that they agree
numerically.

\subsubsection{Properties of the minimax threshold}
We summarize here several known properties of the minimax
threshold (\ref{eq:MinmaxDef}), which provide useful information
about the behavior of SE.

The extremal $F$ achieving the supremum in Eq.~(\ref{eq:MinmaxDef}) is known.
In the case $\gly=+$, it is  a two-point mixture
\begin{eqnarray}
F^+_\eps = (1-\eps)\, \delta_0  + \eps\,\delta_{\mu^+(\eps)}\, .
\end{eqnarray}
In the signed case $\gly=\pm$, it is a three-point symmetric
mixture
\begin{eqnarray}
F^\pm_\eps = (1-\eps)\, \delta_0  + \frac{\eps}{2}\, ( \delta_{\mu^\pm(\eps)} + \delta_{-\mu^\pm(\eps)})\, .
\end{eqnarray}
Precise asymptotic expressions for $\mu^\gly(\eps)$ are available.
In particular, for $\gly\in\{+,\pm\}$,
\begin{eqnarray}
 \mu^\gly(\eps) = \sqrt{2 \log(\eps)} (1 + o(1)) \qquad \mbox{ as }  \eps \goto  0 \, .
\end{eqnarray}
We also know that
\beq \label{eq:Mstar-asymp}
 M^*(\eps; \gly)  = 2 \log(\eps) ( 1+ o(1))  \qquad \mbox{ as }  \eps \goto  0 \, .
\eeq
%
%

\subsubsection{Proof of Theorem \ref{thm:CGSEequiv}}
Combining Theorem \ref{thm:rhhose-rhomm} and Eq.~(\ref{eq:Mstar-asymp}),
we get
\begin{eqnarray}
\rho_{\SE}(\delta;\rho) \sim  \frac{1}{2 \log(\delta)} , \qquad
\delta \goto  0\, .\label{eq:SE_Asymp}
\end{eqnarray}
(correction terms that can be explicitly given).
Now we know rigorously from \cite{DoTa09} that the LP-based
phase transitions satisfy a similar relationship:
\begin{theorem}[Donoho and Tanner \cite{DoTa09}]
For $\gly \in \{ +,\pm \}$
\begin{eqnarray}
\rho_{\CG}(\delta,\gly) \sim \frac{1}{2 \log(\delta)}, \qquad  \delta \goto  0.
\end{eqnarray}
\end{theorem}
Combining now with Lemma \ref{eq:SE_Asymp} we
get Theorem \ref{thm:CGSEequiv}. \qed

%
%
\subsection{Rigorous Asymptotic Optimality of Soft Thresholding}

The discussion in the main text, alluded to the possibility of improving on
soft thresholding. Here we give a more formal discussion. We work
in the situations $\gly \in \{ + , \pm \}$.
Let $\widetilde{\eta}$ denote some arbitrary nonlinearity
with tuning parameter $\lambda$.  (For a concrete
example, think of hard thresholding). We can define the minimax
MSE for this nonlinearity in the natural way
\begin{eqnarray}
\widetilde{M}(\eps;\gly) = \inf_\lambda \sup_{F\in \calF^\gly_\eps}
\E_F \left\{ \widetilde{\eta}( X + Z; \lambda) - X)^2 \right\}  \, ,\label{eq:AltMinmaxDef}.
\end{eqnarray}
there is a corresponding minimax threshold $\widetilde{\lambda}(\eps; \gly)$.
We can deploy the minimax threshold in AMP
by setting $\eps = \rho\delta$ and rescaling the threshold
by the effective noise level $\tau = \sigma/\sqrt{\delta}$:
\begin{eqnarray*}
      \mbox{actual threshold at iteration $t$}
      &=& \widetilde{\lambda}(\eps;\gly) \cdot \tau  \\
      &=& \widetilde{\lambda}(\rho\delta;\gly) \cdot \sigma_t / \sqrt{\delta}.
\end{eqnarray*}
Under state evolution, this is guaranteed to reduce the MSE provided
\[
   \widetilde{M}(\rho\delta;\gly) < \delta .
\]
In that case we get full evolution to zero. It makes sense to define
the minimax phase transition:
\[
    \widetilde{\rho}_{\SE}(\delta ; \gly) =
     \sup\{ \rho\;:\;\; \widetilde{M}(\rho \delta; \gly) < \delta \} ; \qquad \gly \in \{ +, \pm \} .
\]
Whatever be $F$, for $(\delta, \rho)$ with $\rho < \widetilde{\rho}_{\SE}(\delta)$,
SE evolves the formal MSE of $\widetilde{\eta}$  to zero.

It is tempting to hope that some very special nonlinearity
 can do substantially better than soft thresholding.
 At least for the
minimax phase transition, this is not so:
\begin{theorem}  \label{thm:altPT}
Let $\widetilde{\rho}_{\MM}(\delta;\gly)$ be a minimax phase transition computed
under the State Evolution formalism
for the cases $\gly \in \{ +,\pm)$ with some scalar nonlinearity $\widetilde{\eta}$.
Let $ \rho_{\SE}(\delta;\gly)$ be the phase transition calculated in the main text
for soft thresholding with corresponding optimal $\lambda$. Then
for $\gly \in \{ +,\pm \}$
\[
     \lim_{\delta \goto 0}  \frac{\widetilde{\rho}_{\SE}(\delta;\gly)}{\rho_{\SE}(\delta;\gly)} \leq 1.
\]
\end{theorem}

In words, no other nonlinearity can outperform soft thresholding
in the limit of extreme undersampling -- in the sense of minimax
phase transitions.  This is best understood
using a notion from the main text.  We there said that the
parameter space $(\delta,\rho,\lambda,F)$ can be partitioned
into two regions. Region (I) where there zero is the unique fixed point
of the MSE map, and is a stable fixed point; and its complement, Region (II).
Theorem \ref{thm:altPT} says that the range of $\rho$ guaranteeing
membership in Region (I) cannot be dramatically expanded
by using a different nonlinearity.

\subsubsection{Some results on Minimax Risk}
The proof depends on some know results about minimax
MSE, where we are allowed to choose not just the
threshold, but also the nonlinearity.
For $\gly \in \{ +, \pm \}$, define the minimax MSE
\begin{eqnarray}
M^{\star\star}(\eps;\gly) = \inf_{\widetilde{\eta}} \sup_{F\in \calF^\gly_\eps}
\E_F \left\{ \widetilde{\eta}( X + Z) - X)^2 \right\}  \, ,\label{eq:MinmaxMSEDef}
\end{eqnarray}
Here the minimization is over {\it all measurable functions}
$\widetilde{\eta} : \bR \mapsto \bR$. {\it Minimax MSE} was discussed  for the case
$\gly = +$ in \cite{DJHS92} and  for $\gly=\pm$ in \cite{Bickel,DJ94a,DJ94b}.
It is known that
\beq \label{eq:asympMMRisk}
     M^{\star\star}(\eps;\gly) \sim 2 \log(\eps^{-1}). \qquad \eps \goto 0.
\eeq

\subsection{Proof of Theorem \ref {thm:altPT}}

Evidently, any specific nonlinearity cannot do better than
the minimax risk:
\[
     \widetilde{M}^*(\eps) \geq  M^{**}(\eps;\gly).
\]
Consequently, if we put
\[
      \rho^{**} (\delta;\gly) = \sup \{ \rho :\;   M^{**}(\delta\rho;\gly) < \delta  \}
\]
then
\[
       \widetilde{\rho}^{*} (\delta,\gly) \leq   \rho^{\star\star} (\delta,\gly).
\]
From (\ref{eq:asympMMRisk})
and the last two displays we conclude
\[
    \widetilde{\rho}^{*} (\delta;\gly)  \leq \frac{1}{2 \log(1/\delta)}  \sim   \rho_{\SE}(\delta,\gly) , \qquad  \delta \goto 0.
\]
Theorem \ref{thm:altPT} is proven. \qed
%
%
%
%

\subsection{Data Generation}
For a given algorithm with a fully specified parameter vector,
we conduct one phase transition measurement experiment as follows.
We fix a {\it problem suite},
i.e. a matrix ensemble and a coefficient distribution for generating
problem instances $(A,x_0)$.  We also fix a grid of $\delta$ values in $[0,1]$,
typically $30$ values  equispaced between
$0.02$ and  $0.99$. Subordinate to this grid, we consider
a series of $\rho$ values. Two cases arise frequently:
\bitem
\item {\it Focused Search design.} 20 values between
$\rho_{\CG}(\delta;\gly) - 1/10$
and $\rho_{\CG}(\delta;\gly) + 1/10$, where $\rho_{\CG}$ is the theoretically expected phase transition
deriving from combinatorial geometry
(according to case $\gly\in\{+,\pm,\Box\}$).
\item {\it General Search design.} 40 values equispaced between 0 and 1.
\eitem
We then have a (possibly non-cartesian) grid of $\delta,\rho$ values
in parameter space $[0,1]^2$.
At each $(\delta, \rho)$ combination, we will take
$M$ problem instances; in our case $M=20$.
We also fix a measure of success; see below.

Once we specify the problem size $N$, the experiment is now fully specified;
we set $n = \lceil \delta N \rceil$ and $k = \lceil \rho n \rceil$, and generate
$M$ problem instances, and obtain $M$ algorithm outputs $\hat{x}_i$, and
$M$ success indicators $S_i$, $i=1,\dots M$.

A problem instance $(y,A,x_0)$ consists
of  $n \times N$ matrix $A$ from the given matrix ensemble
and a $k$-sparse vector $x_0$ from the given coefficient ensemble.
Then $y = Ax_0$.  The algorithm is called with problem instance $(y,A)$ and
it produces a result  $\hat{x}$. We declare success if
\[
  \frac{\| x_0 - \hat{x} \|_2}{ \| x_0 \|_2 } \leq \mbox{\tt tol} ,
\]
where ${\tt tol}$ is a given parameter; in our case $10^{-4}$;
the variable $S_i$ indicates success on the $i$-th Monte Carlo
realization.
To summarize all $M$ Monte Carlo repetitions,
we set $S = \sum_i S_i$.

The result of such an experiment is a dataset with
tuples $(N,n,k,M,S)$; each tuple giving the results
at one combination $(\rho,\delta)$. The meta-information
describing the experiment is the specification of the
algorithm with all its parameters, the problem suite,
and the success measure with its tolerance.

\subsection{Estimating Phase Transitions}

From such a dataset we find the location of the
 phase transition as follows.
Corresponding to each fixed value of $\delta$ in our grid,
we have a collection of tuples $(N,n,k,M,S)$
with $n/N=\delta$ and varying $k$. Pretending that our
random number generator makes truly independent
random numbers, the result $S$ at one experiment
is binomial $\Bin(\pi,M)$,  where the success probability $\pi \in [0,1]$.
Extensive prior experiments show that this probability
varies from $1$ when $\rho$ is well below $\rho_{\CG}$ to
$0$ when $\rho$ is well above $\rho_{\CG}$.
In short, the success probability
\[
   \pi = \pi(\rho | \delta ; N ) .
\]

We define the {\it finite-$N$
phase transition} as the  value of $\rho$
at which success probability is 50\%:
\[
    \pi(\rho | \delta ; N )  = \frac{1}{2}\;\; \mbox { at }\;\;  \rho = \rho(\delta) .
\]
This notion is well-known in biometrics
where the 50\% point of the dose-response
is called the LD50. (Actually we have the implicit dependence
$ \rho(\delta) \equiv \rho(\delta |N,\mbox{\tt tol} ) $;
the tolerance in the success definition has
a (usually slight) effect, as well as the problem size $N$)

To estimate the phase transition
from data, we model dependence
of success probability on $\rho$ using generalized
linear models (GLMs).  We take a $\delta$-constant
slice of the dataset obtaining triples
$(k,M,S(k,n,N))$, and model $S(k,n,N) \sim \Bin(\pi_k; M)$
where the success probabilities
obeys a generalized linear model with logistic link
\[
      logit(\pi) = a + b \rho
\]
where  $\rho = k/n$; in biometric language, we are modeling
that the dose-response probability, where $\rho$ is the `complexity-dose',
follows a logistic curve.

In terms of the fitted parameters $\hat{a}$,$\hat{b}$,
we have the estimated phase transition
\[
     \hat{\rho}(\delta) = -\hat{a}/\hat{b},
\]
and the estimated transition width is
\[
      \hat{w}(\delta)  = 1/b.
\]
Note that, actually,
\[
     \hat{\rho}(\delta) =  \hat{\rho}(\delta | N, \mbox{\tt tol} ) , \qquad
      \hat{w}(\delta)  = \hat{w}(\delta | N, \mbox{\tt tol} )\, .
\]
We may be able to see the phase transition and its width varying
with $N$ and with the success tolerance.

Because we make only $M$ measurements
in our Monte Carlo experiments, these results are
subject to sampling fluctuations.
Confidence statements can be made for $\hat{\rho}$ using standard
statistical software.
%
%

\subsection{Tuning of Algorithms}
The procedure so far gives us, for each fully-specified
combination of algorithm parameters $\Lambda$ and each problem suite $\cS$,
a dataset $(\Lambda,\cS,\delta,\hat{\rho}(\delta; \Lambda,S))$.
When an algorithm has such parameters,
we can define, for each fixed $\delta$, the
value of the parameters which gives the highest transition:
\[
      \hat{\rho}^{opt}(\delta;\cS) = \max_\Lambda \hat{\rho}(\delta;\Lambda,\cS) ;
\]
with associated optimal parameters $\Lambda^{opt}(\delta; \cS)$.
When the results of the algorithm depend strongly on problem
suite as well, we can also tune to optimize worst-case performance
across suites, getting the minimax transition
\[
  \hat{\rho}^{\MM}(\delta) =  \max_\Lambda \min_\cS \hat{\rho}(\delta;\Lambda,\cS) .
\]
and corresponding minimax parameters $\Lambda^{\MM}(\delta)$.
This procedure was followed in \cite{MaDo09sp} for a wide range of popular algorithms.
Figure 3 of the main text presents the observed minimax transitions.
\begin{figure}[t]
\includegraphics[width=1.1\linewidth]{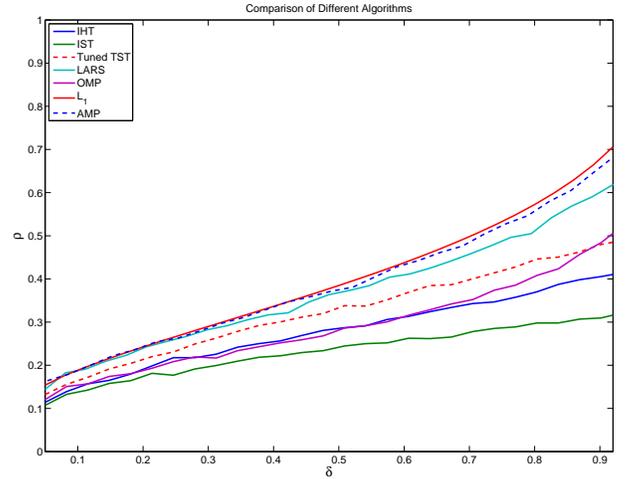}
\caption{Observed Phase Transitions for 6 Algorithms, and $\rho_{\SE}$.
AMP: method introduced in main text.
IST: Iterative Soft Thresholding.
IHT: Iterative Hard Thresholding.
TST: a class of two-stage thresholding algorithms
including subspace pursuit and CoSamp.
OMP: Orthogonal Matching Pursuit.
Note that the $\ell_1$ curve coincides with
the state evolution transition $\rho_{\SE}$,
a theoretical calculation.  The other curves show
empirical results.}
\label{fig:comparePT}
\end{figure}

\subsection{Results: Empirical Phase Transition}
Figure \ref{fig:comparePT} (which is a complete version of Figure 3 in the main text) compares
observed phase transitions of several algorithms including AMP.
We considered what was called in \cite{MaDo09sp} the {\it standard suite},
wit these choices
\bitem
\item Matrix ensemble: Uniform spherical ensemble(USE);
each column of $A$ is drawn uniformly at random from the unit sphere
in $\reals^n$.
\item Coefficient ensemble:
The vector $x_0$ has $k$ nonzeros in random locations,
with constant amplitude of nonzeros.
If $\gly = +$,  $x_0(i)\in\{0,+1\}$; if $\gly\in \{ \pm,\Box\}$,
$x_0(i)\in\{+1,0,-1\}$ (with equiprobable positive and negative entries).
\eitem

For each algorithm we generated an appropriate grid of $(\delta,\rho)$
and created $M=20$ independent problem instances at each gridpoint,
i.e. independent realizations of vector $x$ and measurement matrix $A$.

For AMP we used a focused search design,
focused around $\rho_{\CG}(\delta)$.
To reconstruct $x$, we run $T=1000$ AMP iterations and report the mean square error at the final iteration.
For other algorithms, we used the general search design as described above.
 For more details about observed phase transitions
 we refer the reader  to \cite{MaDo09sp}.

 The calculation of the phase transition curve of AMP
takes around $36$ hours on a single Pentium $4$ processor.

Observed Phase transitions for other coefficient ensembles and matrix ensembles
are discussed below in sections \ref{sec:coefuniv} and \ref{sec:matrixuniv}.

%
%
\subsection{Example of the Interference Heuristic}

In the main text, our motivation of the
 SE formalism used the assumption that the
 mutual access interference term
 MAI${}_t = (A^*A -I) (x^t - x_0)$
is marginally nearly Gaussian -- i.e.
the distribution function of the entries
in the MAI vector is approximately Gaussian.

\begin{figure}[t]
\includegraphics[width=1.1\linewidth]{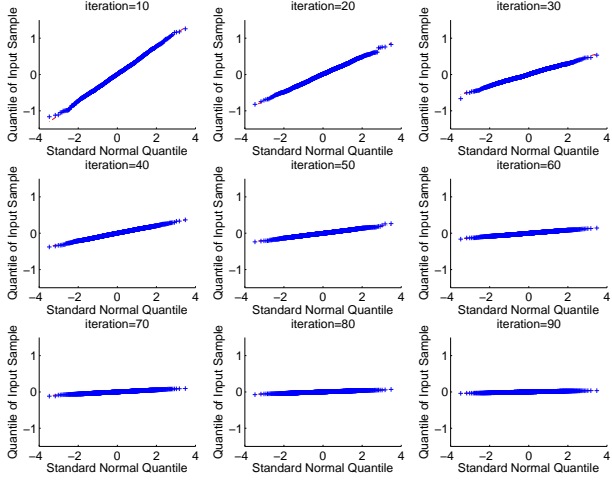}
\caption{QQ Plots tracking marginal distribution of mutual
access interference (MAI).  Panels (a)-(i): iterations $10,20,\dots,90$.
Each panel shows QQ plot of MAI values versus normal distribution in blue,
and in red (mostly obscured) points along a  straight line.  Approximate linearity
indicates approximate normality.  Decreasing slope
with increasing iteration number indicates
decreasing standard deviation as iterations progress. }\label{fig:qqMAI}
\end{figure}

\begin{figure}[t]
\includegraphics[width=1.1\linewidth]{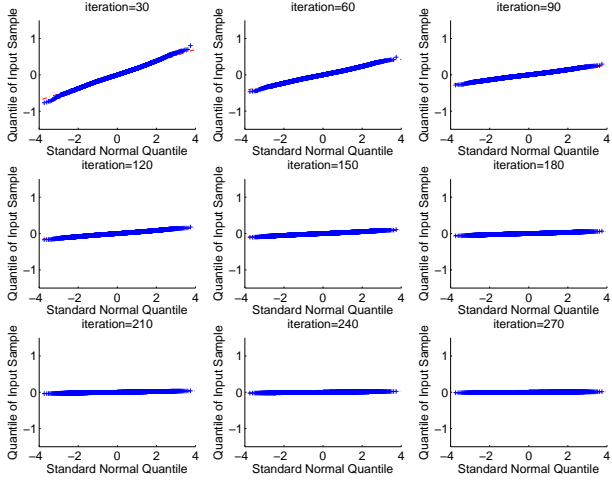}
\caption{QQ Plots tracking marginal distribution of mutual
access interference (MAI).  Matrix Ensemble: partial Fourier.
Panels (a)-(i): iterations 30,60,\dots, 270.
 For other details, see  Fig. \ref{fig:qqMAI}.}\label{fig:qqMAIFourier}
\end{figure}

As we mentioned, this heuristic motivates the
definition of the MSE map. It is easy to prove that the heuristic
is valid at the first iteration; but for the validity of SE,
it must continue to be true at every iteration
until the algorithm stops. Figure \ref{fig:qqMAI}
presents a typical example. In this example we have considered USE matrix ensemble and Rademacher Coefficient ensemble. Also $N$ is set to a small size problem $2000$ and $(\delta,\rho)=(0.9,0.52)$.
The algorithm is
tracked across 90 iterations. Each panel
exhibits a linear trend, indicating
approximate Gaussianity.  The slope
is decreasing with iteration count.
The slope is the square root of the MSE,
and its decrease indicates that the MSE
is evolving towards zero. More interestingly, figure \ref{fig:qqMAIFourier} shows the QQplot of the MAI noise for the partial Fourier matrix ensemble. Coefficients here are again from Rademacher ensemble and $(N,\delta,\rho)=(16384,0.5,0.35)$.

%
%

\subsection{Testing Predictions of State Evolution}
\begin{figure}[t]
\includegraphics[width=1.1\linewidth]{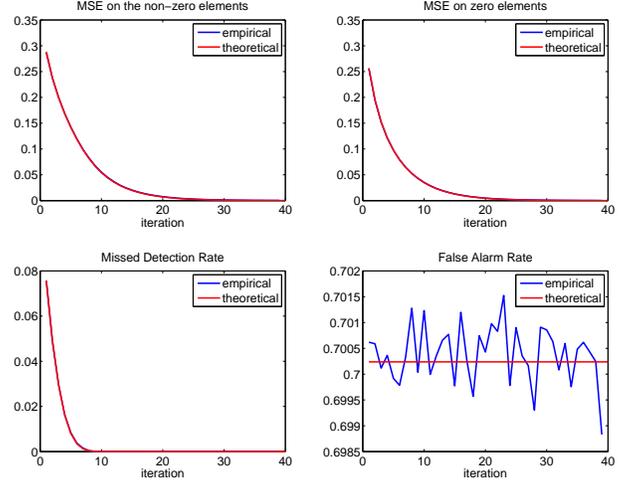}
\caption{Comparison of State Evolution predictions
against observations.  $\rho=.3$, $\delta=.15$.
Panels (a)-(d):  MSENZ, MSE, MDR, FAR.
Curve in red: theoretical prediction. Curve in blue:
mean observable.
Each panel shows the evolution of a specific observable as
iterations progress. Two curves are present in each panel,
however, except for the lower left panel, the blue
curve (empirical data) is obscured by the presence of the red curve.
The two curves are in close agreement in all panels. }\label{fig:TestSE1}
\end{figure}

\begin{figure}[t]
\includegraphics[width=1.1\linewidth]{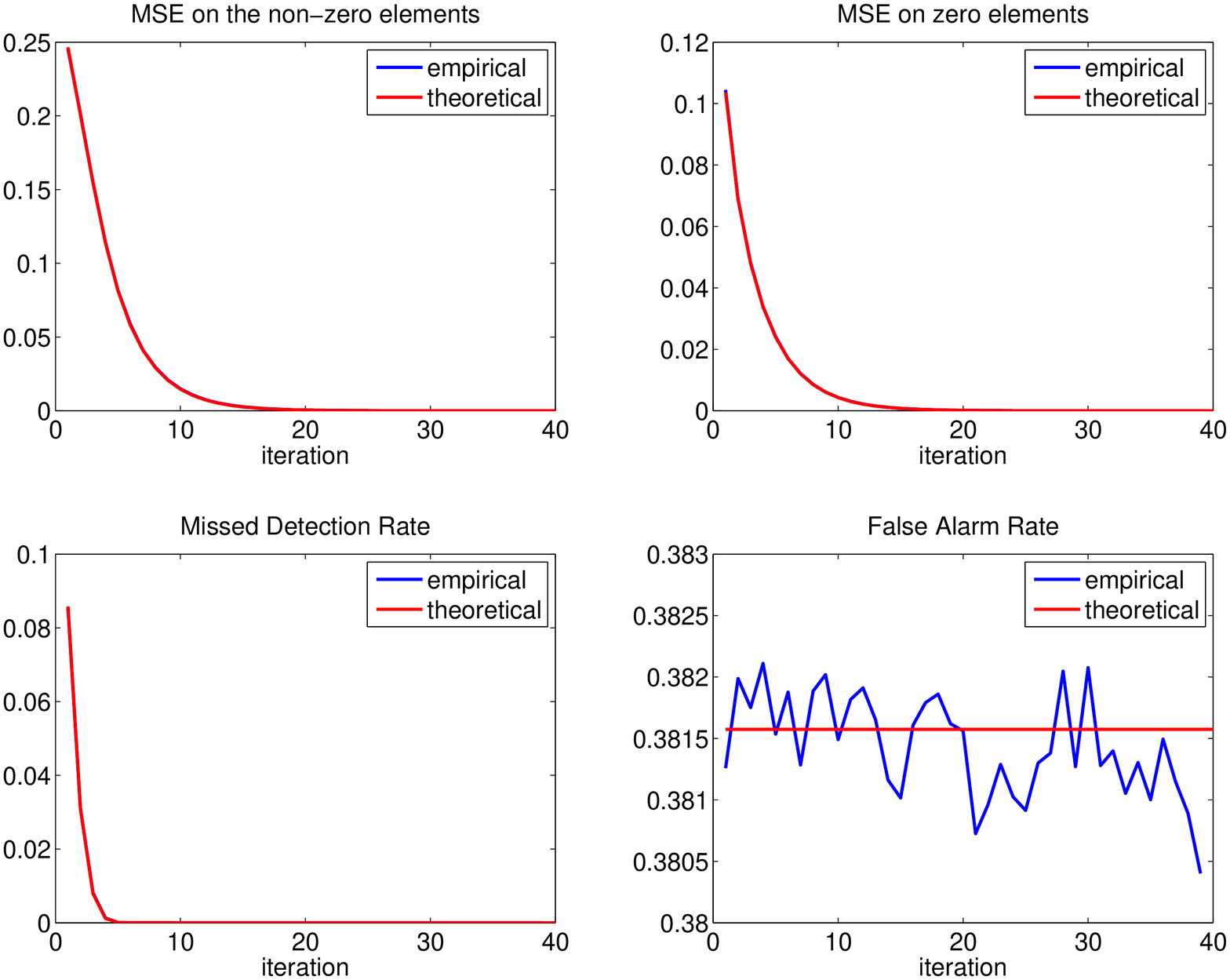}
\caption{Comparison of State Evolution predictions
against observations.  $\rho=0.3$, $\delta=0.15$.
For details, see Figure \ref{fig:TestSE1}.}\label{fig:TestSE2}
\end{figure}
\begin{figure}[t]
\includegraphics[width=1.1\linewidth]{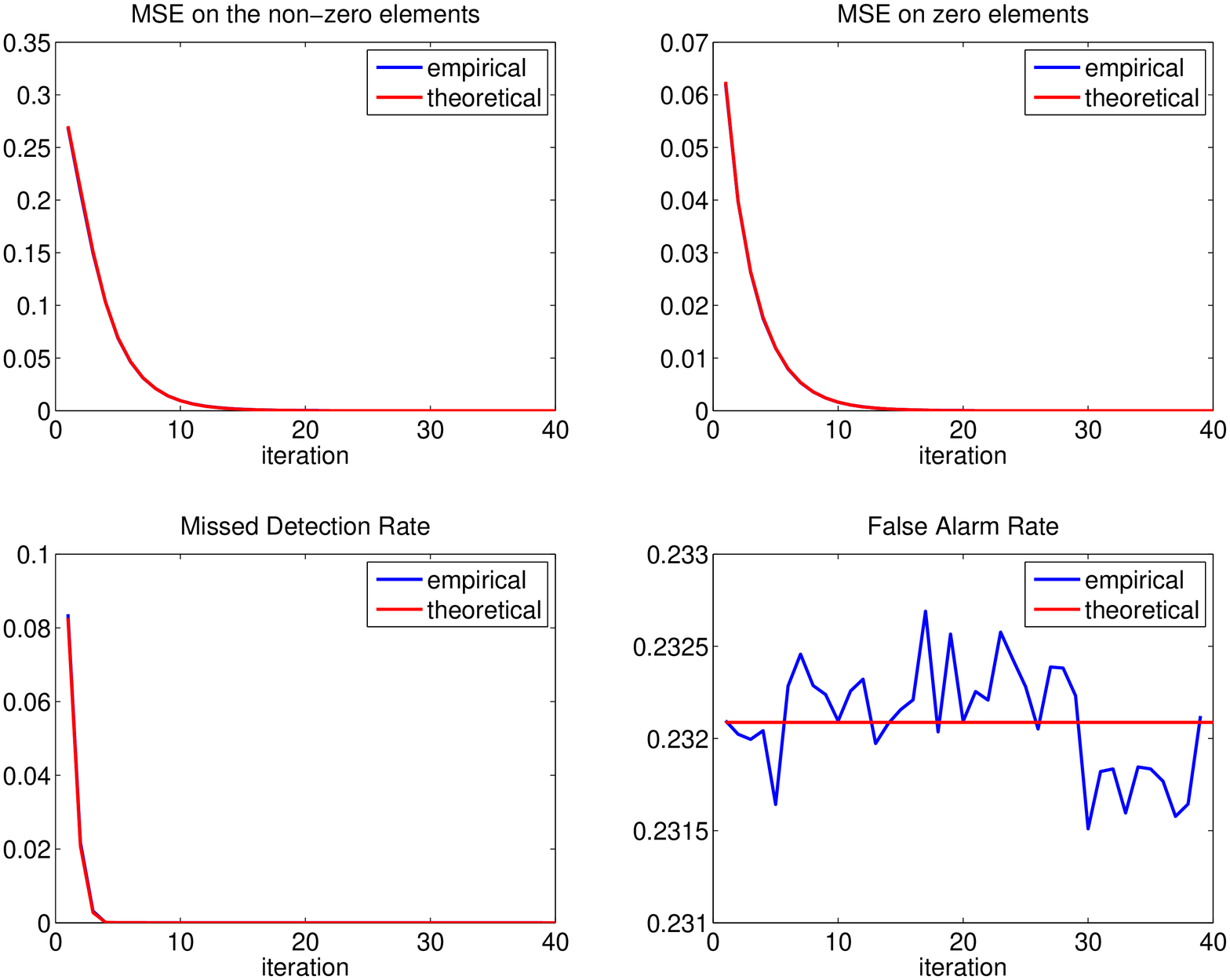}
\caption{Comparison of State Evolution predictions
against observations for  $\rho=0.7$, $\delta=0.36$.
For details, see Figure \ref{fig:TestSE1}.}\label{fig:TestSE3}
\end{figure}
The last section gave an illustration
tracking the actual evolution of the AMP algorithm,
it showed that  the State Evolution
heuristic is qualitatively  correct.

We now consider predictions made by SE
and their quantitative match with empirical observations.
We consider predictions of four observables:
\bitem
\item MSE on zeros and MSE on non-zeros:
\begin{align}
    \mbox{\tt MSEZ}  &=& \expect [ \hat{x}(i)^2 | x_0(i) = 0 ] ,  \nonumber \\
    \mbox{\tt MSENZ}   &=& \expect [ (\hat{x}(i)- x_0(i)) ^2 | x_0(i) \neq 0 ]
\end{align}
\item Missed detection rate and False alarm rate:
\begin{align}
    \mbox{\tt MDR}  &=& \prob [ \hat{x}(i)  = 0      | x_0(i) \neq  0 ], \nonumber \\
    \mbox{\tt FAR}   &=& \prob [ \hat{x}(i) \neq 0  | x_0(i) =  0 ]
\end{align}
\eitem

We illustrate the calculation of {\tt MDR}. Other quantities are computed similarly.
Let $\eps = \delta \rho$, and suppose that  entries in $x_0(i)$  are either
$0$, $1$, or $-1$, with  $\prob\{ x_0(i) = \pm1 \} = \eps/2$.
Then, with $Z \sim N(0,1)$,
\begin{eqnarray}
   \prob [ \hat{x}(i)  = 0 | x_0(i) \neq  0 ]  &=& \prob [ \eta( 1 + \frac{\sigma}{\sqrt{\delta}} Z )  \neq 0 ] \nonumber \\
       &=&  \prob [     1 + \frac{\sigma}{\sqrt{\delta}}Z  \not \in (-\lambda\sigma,\lambda\sigma ) ] \nonumber \\
       &=&  \prob [     Z  \not \in (a,b ) ]
\end{eqnarray}
with $a = ((-\lambda -1/\sigma) \cdot \sqrt{\delta}$, $b = (\lambda-1/\sigma) \cdot \sqrt{\delta}$.

In short, the calculation merely requires classical properties of the normal
distribution. The three other quantities simply require other similar properties
of the normal.
As discussed in the main text, SE evolution makes an iteration-by-iteration
prediction of $\sigma_t$; in order to calculate predictions
of {\tt MDR}, {\tt FAR}, {\tt MSENZ} and {\tt MSEZ}, the
parameters $\eps$ and $\lambda$ are also needed.

We compared the state evolution predictions with the actual values
by a Monte Carlo experiment.  We chose these triples
$(\delta,\rho,N)$: $(0.3,0.15,5000)$, $(0.5,0.2,4000)$, $(0.7,0.36,3000)$.
We again used the standard problem suite (USE matrix and unit amplitude nonzero).
At each combination of $(\delta,\rho,N)$,
we generated $M=200$ random problem instances from the standard problem suite,
and ran the AMP algorithm for a fixed number of iterations.
We computed the observables at each iteration. For example,
the empirical missed detection rate is estimated by
\[
    {\tt eMDR}(t) = \frac{ \#\{ i :  x^t(i) = 0  \;\;\mbox{ and }\;\;
 x_0(i) \neq 0 \}}{ \#\{ i : x_0(i) \neq 0 \}}.
\]
We averaged the observable trajectories across the $M$ Monte Carlo
realizations, producing empirical averages.

 The results for the three cases are presented in
 Figures \ref{fig:TestSE1}, \ref{fig:TestSE2}, \ref{fig:TestSE3}.
 Shown on the display are curves indicating both the
 theoretical prediction and the empirical averages.
 In the case of the upper row and the lower left panel,
 the two curves are so close that one cannot easily
 tell that two curves are, in fact, being displayed.
%
%
\subsection{Coefficient Universality}
\label{sec:coefuniv}

\begin{figure}[t]
\includegraphics[width=1.1\linewidth]{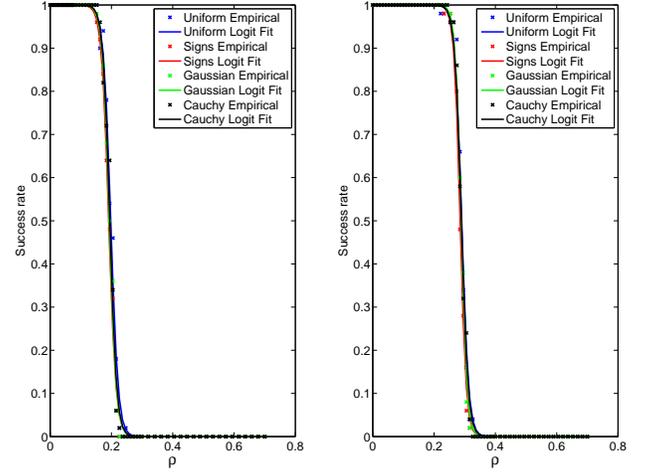}
\caption{Comparison of Failure probabilities
for different ensembles.  In the left window, $\delta=0.10$ and in the right window $\delta=0.3$.
Red: unit-amplitude coefficients.
Blue: uniform $[-1,1]$.
Green: Gaussian.
Black: Cauchy.
Points: observed failure fractions
Curves: Logistic fit.}\label{fig:Coeff1}
\end{figure}

SE displays  invariance of the
evolution results with respect to the coefficient distribution
of the nonzeros. What happens in practice?

We studied invariance of AMP results
as we varied the distributions of the nonzeros in $x_0$.
We consider the problem $\gly=\pm$ and used the following
distributions for the non-zero entries of $x_0$:
\bitem
\item Uniform in $[-1,+1]$;
\item Radamacher (uniform in $\{+1,-1\}$);
\item Gaussian;
\item Cauchy.
\eitem
In this study,
 $N=2000$, and we considered $\delta = 0.1$, $0.3$.
For each value of $\delta$ we considered
$20$ equispaced values of $\rho$ in the interval
$[\rho_{\CG}(\delta;\pm)-1/10,\rho_{\CG}(\delta;\pm)+1/10]$, running each time
$T=1000$ AMP iterations.
Data are presented, respectively, in
Figures \ref{fig:Coeff1}.

Each plot displays the fraction of success  $(S/M)$
as a function of $\rho$ and a fitted success probability i.e. in terms
of success probabilities,  the curves display  $ \pi(\rho)$.
In each case 4 curves and 4 sets of data points are displayed,
corresponding to the 4 ensembles. The four datasets
are visually quite similar, and it is apparent that
indeed a considerable degree of invariance
is present.

%
%
\subsection{Matrix Universality}
\label{sec:matrixuniv}

The Discussion section in the  main text
referred to evidence that our results
are not limited to the Gaussian distribution.

\begin{figure}[t]
\includegraphics[width=1.1\linewidth]{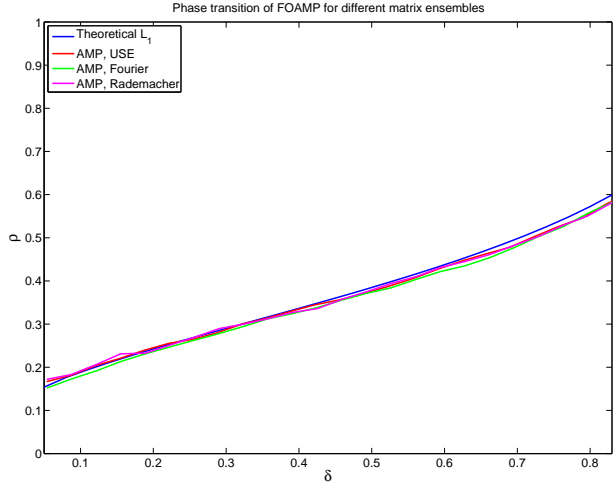}
\caption{Observed Phase Transitions at different matrix ensembles.   Case $\gly = \pm$.
Red:  Uniform Spherical Ensemble (Gaussian with normalize column lengths).
Magenta: Rademacher ($\pm 1$ equiprobable).
Green: partial Fourier.
Blue:  $\rho_{\ell_1}$.}\label{fig:Matrix1}
\end{figure}

We conducted a study of AMP where
everything was the same as in Figure 1 above,
however, the matrix ensemble  could change.
We considered three such ensembles:
USE (columns iid uniformly distributed on
the unit sphere), Rademacher (random entries iid $\pm 1$ equiprobable),
and Partial Fourier, (randomly select $n$ rows from $N \times N$ fourier matrix.)
We only considered the case $\gly = \pm$.
Results are shown in Fig.~\ref{fig:Matrix1}, and compared
to the theoretical phase transition for $\ell_1$.

%
%

\subsection{Timing Results}
In actual applications, AMP runs rapidly.

We first describe a study comparing AMP
to the LARS algorithm \cite{EfHaJoTi04}.  LARS
is appropriate for comparison because, among the iterative
algorithms previously proposed, its phase
transition is closest to the $\ell_1$ transition.
So it comes closest to duplicating
the AMP sparsity-undersampling
tradeoff.

Each algorithm proceeds iteratively and needs
a stopping rule. In both cases, we stopped calculations when
the relative fidelity measure exceeded $0.999$, ie when
$\|y-Ax^t\|_2/ \|y\|_2 <0.001 $.

In our study, we used the partial Fourier matrix ensemble
with unit amplitude for nonzero entries in the signal $x_0$.
We  considered a range of problem sizes $(N,n,k)$ and
in each case  averaged timing results
over $M= 20$ problem instances.  Table \ref{tbl:cmpLARS}
presents timing results.

In all situations studied,  AMP is substantially faster
than LARS. There are a few very sparse
situations -- i.e. where $k$ is in the tens or few hundreds -- where
LARS performs relatively well, losing the race by less
than a factor 3.  However, as the complexity of
the objects increases, so that $k$ is several hundred or even
one thousand, LARS is beaten by factors of 10 or even more.

(For very large $k$, AMP has a decisive advantage. When the matrix $A$
is dense, LARS requires at least $ c_1 \cdot k \cdot n \cdot N $ operations,
while AMP requires at most $ c_2 \cdot n \cdot N $ operations. Here
$c_2 = \log( (\expect X^2) / \sigma^2_T)/b$ is a bound on the number
of iterations, and $(\expect X^2) / \sigma^2_T$ is the relative improvement
in MSE in $T$ iterations. Hence in terms of flops we have
\[
     \frac{{\sf flops}({\rm LARS})}{{\sf flops}({\rm AMP})} \geq \frac{ k b(\delta,\rho)}{\log( (\expect X^2) / \sigma^2_T)}\, .
\]
This  logarithmic dependence of the denominator is very weak,
and very roughly this ratio scales directly with $k$.)

\begin{table}
\caption{Timing Comparison of AMP and LARS. Average
Times in CPU seconds.} 
\centering 
\vspace{.5cm}
\begin{tabular}{|l|l|l|l|l|}
  \hline
  $N$    & $n$ & $k$ & AMP & LARS \\
  \hline
  4096 & 820 & 120 & 0.19 & 0.7 \\
  \hline
  8192 & 1640 & 240 & 0.34 & 3.45 \\
  \hline
  16384 & 3280 & 480 & 0.72 & 19.45 \\
  \hline
  32768 & 1640 & 160 & 2.41 & 7.28 \\
  \hline
  16384 & 820 & 80 & 1.32 & 1.51 \\
  \hline
  8192 & 820 & 110 & 0.61 & 1.91 \\
  \hline
  16384 & 1640 & 220 & 1.1 & 5.5 \\
  \hline
  32768 & 3280 & 440 & 2.31 & 23.5 \\
  \hline
  4096 & 1640 & 270 & 0.12 & 1.22 \\
  \hline
  8192 & 3280 & 540 & 0.22 & 5.45 \\
  \hline
  16384 & 6560 & 1080 & 0.45 & 27.3 \\
  \hline
  32768 & 1640 & 220 & 6.95 & 17.53 \\
  \hline
\end{tabular}
\label{timing}
\label{tbl:cmpLARS}
\end{table}

We also studied AMP's ability to
solve very large problems.

We conducted a series of trials with increasing $N$
in a case where $A$ and $A^*$ can be applied
rapidly, without using ordinary matrix storage
and matrix operations; specifically,
the partial Fourier ensemble.
For nonzeros of  the signal $x_0$.
we  chose unit amplitude nonzeros.

We considered the fixed choice
$(\delta,\rho) = (1/6,1/8)$
and $N$ ranging from $1K$
to ($K=1024$)     to $256K$ in powers of $2$.
At each signal length $N$ we
 generated $M=10$ random problem instances
 and measured CPU times (on a single Pentium 4 processor) and iteration counts
 for AMP in each instance.
 We considered four stopping rules, based on
 MSE $\sigma^2$, $ \sigma^2/2$, $\sigma^2/4$,
 and $\sigma^2/8$,
 where $\sigma^2 = 12 \cdot 10^{-5}$.
We then averaged timing results
over the $M= 10$ randomly generated problem instances

\begin{figure}[t]
\begin{center}
\includegraphics[width=3.5in]{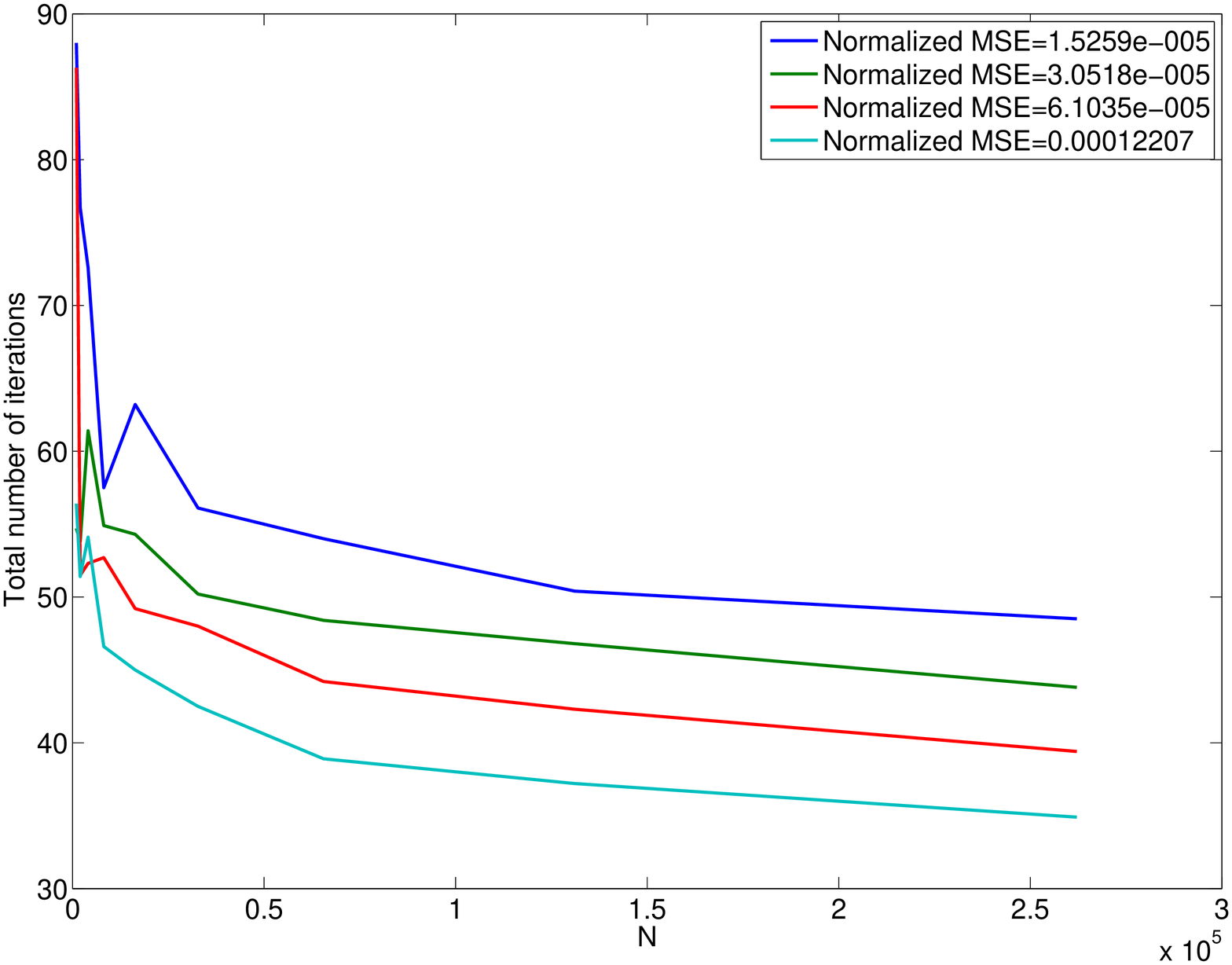}
\caption{Iteration Counts versus Signal Length $N$.
Different curves show results for different stopping rules.
Horizontal axis: signal length $N$.
Vertical axis: Number of iterations, $T$.
Blue, Green, Red, Aqua curves depict
results when stopping thresholds are set at  $12 \cdot 10^{-5} 2^{4-\ell}$,
with $\ell = 0,1,2,3$ Each doubling of accuracy costs about 5 iterations.} \label{fig:Timing1}
\end{center}
\end{figure}

 Figure \ref{fig:Timing1} presents the number of iterations
 as a function of the problem size and accuracy level.
 According to the SE formalism, this should be a constant
 independent of $N$ at each fixed $(\delta,\rho)$
 and we see indeed that this is the case for AMP:
 the number of iterations is close to constant for
all large $N$.
Also according to the SE formalism, each additional
iteration produces a proportional reduction in formal MSE,
and indeed in practice each increment of $5$ AMP iterations
reduces the actual MSE by about half.

\begin{figure}[t]
\begin{center}
\includegraphics[width=3.5in]{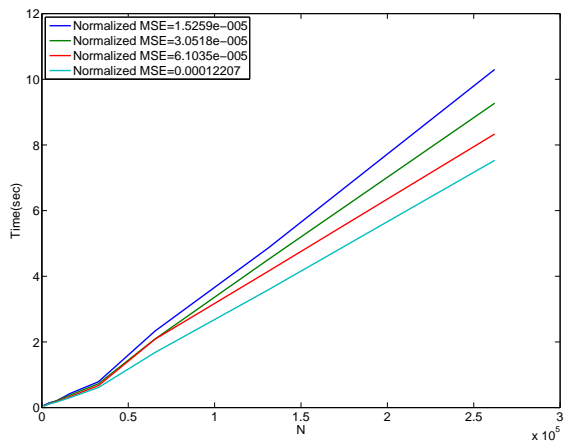}
\caption{CPU Time Scaling with $N$ .
Different curves show results for different stopping rules.
Horizontal axis: signal length $N$.
Vertical axis: CPU time(seconds).
Blue, Green, Red, Aqua curves depict
results when stopping thresholds are set at  $12 \cdot 10^{-5} 2^{4-\ell}$,
with $\ell = 0,1,2,3$} \label{fig:Timing2}
\end{center}
\end{figure}

Figure \ref{fig:Timing2} presents CPU time
 as a function of the problem size and accuracy level.
 Since we are using the partial Fourier ensemble,
the cost of applying $A$ and $A^*$ is proportional to  $N \log(N)$;
this is much less
 than what we would expect for
 the cost of applying a general dense matrix.
 We see that indeed AMP execution time scales very favorably with $N$
 in this case -- to the eye, the timing seems practically linear with $N$.
The timing results show that each doubling of $N$
produces essentially a doubling of execution time.
iteration produces a proportional reduction in formal MSE,
and indeed in practice each increment of $5$ AMP iterations
reduces the MSE by about half.
Each doubling of accuracy costs about 30\% more computation time.

\end{document}